\documentclass[referee, natbib]{mn2e}
\usepackage{amsmath}
\usepackage{amssymb}
\usepackage{graphicx}
\usepackage{color}
\newcommand{\ms}{m_{\star}}

\newcommand{\me}{{\rm M}_\oplus}
\newcommand{\rearth}{{\rm R}_\oplus}
\newcommand{\msun}{{\rm M}_\odot}
\newcommand{\rcav}{{\rm r}_{\rm cav}}
\newcommand{\rc}{{\rm r}_{\rm crit}}

\newcommand{\ai}{a_{\rm i}}
\newcommand{\ei}{e_{\rm i}}
\newcommand{\tei}{t_{e,{\rm i}}}
\newcommand{\tai}{t_{a,{\rm i}}}
\newcommand{\tmi}{t_{m,{\rm i}}}

\newcommand{\Rone}{\left< \mathcal{R}_{\rm 1} \right> }
\newcommand{\Rtwo}{\left< \mathcal{R}_{\rm 2} \right> }
\newcommand{\RDres}{\left< \mathcal{R}_{\rm D}^{\rm res} \right>}
\newcommand{\RDsec}{\left< \mathcal{R}_{\rm D}^{\rm sec} \right>}

\newcommand{\RE}{\left< \mathcal{R}_{\rm E} \right>}
\newcommand{\RI}{\left< \mathcal{R}_{\rm I} \right>}
\begin{document}

\title[First--order mean motion resonances in two--planet
systems]{First--order mean motion resonances in two--planet systems:
  general analysis and observed systems}

\author[C. Terquem, J. C. B. Papaloizou] { Caroline
  Terquem$^{1,2}$\thanks{E-mail: caroline.terquem@physics.ox.ac.uk,
    J.C.B.Papaloizou@damtp.cam.ac.uk} and John C.
  B. Papaloizou$^{3}$\footnotemark[1] 
  \\
  $^{1}$ Physics Department,
  University of Oxford, Keble Road, Oxford OX1 3RH, UK \\
  $^{2}$ Institut d'Astrophysique de Paris, UPMC Univ Paris 06, CNRS,
  UMR7095, 98 bis bd Arago, F-75014, Paris, France \\
  $^{3}$ DAMTP, University of Cambridge, Wilberforce Road, Cambridge
  CB3 0WA, UK}

%\date{2018}

%\pagerange{\pageref{firstpage}-\pageref{lastpage}} \pubyear{2018}

\maketitle

\label{firstpage}

%
%===========================================================
%

\begin{abstract}
  This paper focuses on two--planet systems in a first--order
  $(q+1):q$ mean motion resonance and undergoing type--I migration in
  a disc.  We present a detailed analysis of the resonance valid for
  any value of $q$.  Expressions for the equilibrium eccentricities,
  mean motions and departure from exact resonance are derived in the
  case of smooth convergent migration. We show that this departure,
  not assumed to be small, is such that period ratio normally exceeds,
  but can also be less than, $ (q+1)/q.$ Departure from exact
  resonance as a function of time for systems starting in resonance
  and undergoing divergent migration is also calculated.  We discuss
  observed systems in which two low mass planets are close to a
  first--order resonance.  We argue that the data are consistent with
  only a small fraction of the systems having been captured in
  resonance.  Furthermore, when capture does happen, it is not in
  general during smooth convergent migration through the disc but
  after the planets reach the disc inner parts.  We show that although
  resonances may be disrupted when the inner planet enters a central
  cavity, this alone cannot explain the spread of observed
  separations.  Disruption is found to result in either the system
  moving interior to the resonance by a few percent, or attaining
  another resonance.  We postulate two populations of low mass
  planets: a small one for which extensive smooth migration has
  occurred, and a larger one that formed approximately {\em in--situ}
  with very limited migration.
\end{abstract}

\begin{keywords}
celestial mechanics -- planetary systems -- planetary systems:
formation -- planetary systems: protoplanetary discs -- planets and
satellites: general
\end{keywords}

%
%===========================================================
%

\section{Introduction}

The orbital architecture of extrasolar planetary systems has been the
focus of many studies since Lissauer et al. (2011) published the first
statistical analysis of {\em Kepler} multiplanet systems based on the
first four months of mission data.  They reported that most of the
systems were not in or close to mean motion resonances (MMRs), but
that at the same time there was a significant excess of planet pairs
near MMRs.  These results were later confirmed by Fabrycky et
al. (2014) using the first six quarters of {\em Kepler} data, who in
addition pointed out that planet pairs near MMRs tend to be
preferentially {\em wide} of exact resonance.

Because of observational bias, the planets detected by {\em Kepler}
are on short period orbits.  They have either formed further away and
migrated over a large distance down to the disc inner parts, or
undergone only modest convergent migration, possibly forming {\em
  in--situ}.  If the outer planet is the more massive migration
usually leads to resonant capture, either while the planets migrate
through the disc, or after they reach a cavity interior to the disc.
In the former case the commensurability is expected to be maintained
while the planets continue to migrate.

Such a scenario leads to a probability of capture which is much higher
than indicated by the data (Izidoro et al. 2017), although resonances
may be overstable and therefore not permanent when the forced
eccentricities are large enough (Goldreich \& Schlichting~2014, Hands
\& Alexander~2018).  {\em In--situ} formation leads to systems which
are not preferentially in resonances (Hansen \& Murray~2013).
Petrovich et al. (2013) note that two planet systems that appear for
the most part to be just wide of resonance can be formed {\em
  in--situ} starting from a non resonant pair by continuously
increasing their masses until a resonant interaction starts to
occur. However, final masses significantly exceed those of
super--Earths.
% and orbital eccentricities are large.

A large number of the studies published so far have assumed that
planets migrate through the disc, capturing each other in resonances,
and have then tried to identify mechanisms able to disrupt resonances.
Small offsets exterior to exact MMRs are a general outcome of
dissipative processes that preserve angular momentum, such as orbital
circularization through interaction with the central star (Papaloizou
\& Terquem~2010, Papaloizou~2011, Lithwick \& Wu~2012, Delisle et
al. ~2012, Batygin \& Morbidelli ~2013).  However, they usually move
the system away from exact resonance by a few percent only.  It has
been proposed that more significant departures may result from
turbulent fluctuations in the disc (Adams et al. 2008, Rein 2012),
interaction between a planet and the wake of a companion (Baruteau \&
Papaloizou 2013) or interaction between the planets and planetesimals
after disc dissipation (Chatterjee \& Ford~2015).  However, as will be
discussed in section~\ref{sec:discussion} of this paper, it is not
clear that these models are able to give a complete explanation of the
data.

Usually, studies of resonances in multiple planet systems use data
related to all multiple systems, without consideration for the number
of planets in each system.  However, it has been pointed out that,
although two--planet systems near resonance could be part of a smooth
distribution of period ratios, the probability of near resonant chains
to be the result of randomness is lower in higher--multiplicity
systems (Fabrycky et al. 2014).  Therefore, it may be that migration
plays a more important role in shaping systems with more than two
planets.  For this reason, we focus here on systems with only two
planets and which are near MMRs.  We do not include adjacent pairs of
planets from higher--multiplicity systems, unless they are clearly too
far away from the other planets in the system to interact with them.
That way, the only interactions in the systems we study are consistent
with being only between the two planets themselves and the planets and
the disc.  We also focus on planets with masses low enough that they
are in the regime of type~I migration.  These restrictions enable us
to better define the conditions in which the systems we study have
evolved, and remove a number of parameters that could affect our
conclusions.  We also restrict our study to first--order MMRs as these
are the resonances in which low--mass planets are most easily captured
during migration (Papaloizou \& Szuszkiewicz~2005).  The second order
5:3 MMR will however also be considered when discussing observations.

Numerous analyses of first--order MMRs for planets subject to
eccentricity damping and/or disc torques that reduce their angular
momentum have been carried out (e.g., Papaloizou \& Terquem~2010,
Papaloizou 2011, Lithwick \& Wu 2012, Batygin \& Morbidelli 2013,
Goldreich \& Schlichting 2014).  In the first part of this paper, we
extend these studies.  In section~\ref{sec:theory}, we give the
equations that govern a first--order MMR to first order in
eccentricities, and give expressions for the eccentricity damping and
orbital migration timescales.  In section~\ref{sec:evolution}, we
calculate the equilibrium values of the eccentricities and departure
from exact resonance in the case of convergent migration.  This
departure is not assumed to be small and it is shown that in some
cases it can lead to the system being interior to as well as wide of
exact resonance. An expression for the departure from exact resonance
as a function of time for systems starting in resonance and undergoing
divergent migration is also derived.  Such an expression has been
obtained previously allowing for orbital circularization and small
times $ t.$ Here, we extend the treatment to include migration torques
and consider larger values of $t,$ so allowing for more extensive
divergence.  In section~\ref{sec:numeric}, we solve Lagrange's
planetary equations numerically and compare the results with those of
the analysis.

In the second part of the paper, we discuss observed systems.  In
section~\ref{sec:observations}, we discuss the data for two planet
systems close to MMR and show that, in the majority of cases,
extensive convergent migration through a smooth disc with
corresponding formation and maintenance of a MMR cannot have happened.
In section~\ref{sec:cavity}, we investigate the evolution of the
system when the inner planet enters a cavity interior to the disc and
consider departures from commensurability that may be
produced. Finally, in section~\ref{sec:discussion}, we summarize and
discuss our results.

%
%===========================================================
%

\section{Equations governing a first--order mean motion resonance}
\label{sec:theory}

In this section, we consider two planets in a first--order MMR, write
the disturbing function to first order in eccentricities, Lagrange's
planetary equations that give the rate of change of the orbital
elements, and include migration and eccentricity damping. 

\subsection{Disturbing function}
\label{sec:df}

We consider two planets of masses $m_1$ and $m_2$ orbiting a star of
mass $\ms$.  The subscripts '1' and '2' refer to the inner and outer
planets, respectively.  The orbital elements $\lambda_{\rm i}$,
$a_{\rm i}$, $e_{\rm i}$, $n_{\rm i}$ and $\varpi_{\rm i}$, with
${\rm i}=1,2$ denote the mean longitude, semi--major axis,
eccentricity, mean motion and longitude of pericenter of the planet of
mass $m_i$.  We suppose that the two planets are close to or in a
$(q+1):q$ mean motion commensurability, i.e.  $n_1/n_2$ is close or
equal to $(q+1)/q$, where $q \ge 1$ is an integer.  The dynamics is
therefore dominated by the resonant and secular terms in the
disturbing function, since all the other terms are short--period and
average out to zero over the orbital periods.

The perturbing functions for the inner and outer planets can
be written under the form (Murray \& Dermott 1999):

\begin{align}
  \Rone & = \frac{G m_2}{a_2} \left( \RDsec + \RDres +\alpha \RE \right),
  \label{Rpli} \\
  \Rtwo & =  \frac{G m_1}{a_2} \left( \RDsec + \RDres + \frac{1}{\alpha^2} 
    \RI \right), \label{Rplo}
\end{align}

\noindent where $G$ is the constant of gravitation,
$\alpha \equiv a_1 / a_2$, $\RDsec$ and $\RDres$ are the secular and
resonant contributions to the direct part of the disturbing function,
respectively, $\RE$ is the contribution of the indirect part due to an
external perturber and $\RI$ is the contribution of the indirect part
due to an internal perturber.  Note that the latter are resonant
contributions, there is no secular contribution to $\RE$ and $\RI$.
The brackets indicate that the quantities are time--averaged.

We assume small eccentricities, and expand the perturbing functions in
the orbital elements to first order in $e_1$ and $e_2$ (Murray \&
Dermott 1999, p.~329):
\begin{align}
 \RDsec & =  
  0 , \label{RDsec} \\ 
  \RDres & =  e_1 f_1 \cos \phi_1 + e_2 f_2 \cos \phi_2 , 
  \label{RDres} \\
  \RE & =   -2 e_2 \cos \phi_2 \; \delta_{q,1}, \label{RE} \\
  \RI & =   - \frac{1}{2} e_2 \cos \phi_2 \; \delta_{q,1} , \label{RI} 
\end{align}
\noindent where $\delta_{q,1} $ is the usual Kronecker symbol.  
%Note that, for $q=1$, $2 \alpha = 1/(2 \alpha^2)$.  
The coefficients $f_1$ and $f_2$ are given by:
\begin{align}
f_1 & =  - \frac{1}{2} \left[ 2(q+1)+ \alpha \frac{{\rm d}}{{\rm d} \alpha}
      \right] b_{1/2}^{(q+1)} ( \alpha) , \label{f1} \\
f_2 & =   \frac{1}{2} \left[ 2q+1+ \alpha \frac{{\rm d}}{{\rm d} \alpha}
      \right] b_{1/2}^{(q)} ( \alpha), \label{f2}
\end{align}
where $b_{1/2}^{(j)}$ is the Laplace coefficient:
\begin{equation}
  b_{1/2}^{(j)}(\alpha) = \frac{1}{\pi}\int_0^{2\pi}
  \frac{\cos(j\psi)}{\left(1-2\alpha\cos\psi+\alpha^2\right)^{1/2}} \; 
  \text{d}\psi.
\end{equation}
The resonant angles $\phi_1$ and $\phi_2$ are defined by:
\begin{align}
\phi_1 & =  (q+1) \lambda_2- q \lambda_1 -\varpi_1, \label{phi1} \\
\phi_2 & =   (q+1) \lambda_2- q \lambda_1 -\varpi_2. \label{phi2} 
\end{align}

\subsection{Lagrange's planetary equations}
\label{sec:lpe}

When the perturbing function is expanded to first order in the
eccentricities, Lagrange equations can be written as
follows:

\begin{align}
  \dot{n}_{\rm i} & =   -\frac{3}{ a_{\rm i}^2}
  \frac{\partial \left< \mathcal{R}_{\rm i} \right>}{\partial
                           \lambda_{\rm i}}, \label{dndt} \\
  \dot{e}_{\rm i} & =   \frac{-1}{n_{\rm i} a_{\rm i}^2 e_{\rm i}}
  \frac{\partial \left< \mathcal{R}_{\rm i} \right>}{\partial \varpi_{\rm i}}, \\
  \dot{\varpi}_{\rm i} & =   \frac{1}{n_{\rm i} a_{\rm i}^2
                                e_{\rm i}}
   \frac{\partial \left< \mathcal{R}_{\rm i} \right>}{\partial e_{\rm i}}, \\
  \dot{\lambda}_{\rm i} & =   n_{\rm i} + 
  \frac{1}{n_{\rm i} a_{\rm i}^2} \left(
    -2 a_{\rm i} \frac{\partial  \left<\mathcal{R}_{\rm i} \right>}{\partial a_{\rm i}}
    +\frac{e_{\rm i}}{2 }
    \frac{\partial \left< \mathcal{R}_{\rm i} \right>}{\partial e_{\rm i}}  \right) , 
  \label{dlambdadt} 
\end{align}

\noindent where ${\rm i}=1,2$.  
 
Equations~(\ref{dndt})--(\ref{dlambdadt}) yield the following
first--order ordinary differential equations for the seven variables
$n_1$, $n_2$, $e_1$, $e_2$, $\varpi_1$, $\varpi_2$ and
$\sigma=(q+1) \lambda_2 - q \lambda_1$, to which we add the two
equations that give the resonant angles $\phi_1$ and $\phi_2$:

\begin{align}
  \dot{n}_1 & = -3q n_1^2  \frac{\alpha m_2}{\ms} \left( e_1 f_1 \sin
                     \phi_1  + e_2 f^{\prime}_2 \sin \phi_2 \right),  \label{dn1dt} \\
  \dot{n}_2 & = 3(q+1) n_2^2 \frac{m_1}{\ms} \left( e_1 f_1 \sin
                     \phi_1  + e_2 f^{\prime}_2 \sin \phi_2  \right), \label{dn2dt} \\
  \dot{e}_1 & = -n_1 \frac{\alpha m_2}{\ms}  f_1 \sin
                    \phi_1, \label{de1dt} \\
  \dot{e}_2 & = -n_2 \frac{m_1}{\ms}   f^{\prime}_2 
                    \sin \phi_2, \label{de2dt} \\
\dot{\varpi}_1 & = n_1 \frac{\alpha m_2}{\ms} 
                        \frac{1}{e_1} f_1 \cos \phi_1
                 , \label{dvarpi1dt}  \\
 \dot{\varpi}_2 & = n_2 \frac{m_1}{\ms} 
                        \frac{1}{e_2}  f^{\prime}_2
                     \cos \phi_2
                  , \label{dvarpi2dt}  \\
\dot{\sigma} & = (q+1) n_2 - q n_1 , \label{dsigmadt} \\
\phi_1 & =\sigma - \varpi_1 , \label{phi1bis}\\
\phi_2 & = \sigma - \varpi_2. \label{phi2bis}
\end{align}

\noindent Here we have used the fact that, for $q=1$,
$2 \alpha = 1/(2 \alpha^2)$, and we have defined
$ f^{\prime}_2 \equiv f_2 - 2 \alpha \delta_{q,1} .$ In writing
equation~(\ref{dsigmadt}), we have retained only the leading terms,
e.g. the zeroth order term in eccentricities.  When $q=1$, these
equations are the same as those of Goldreich \& Schlichting (2014).
%After solving these equations for the seven variables, the resonant
%angles are calculated from $\phi_1=\sigma - \varpi_1$ and
%$\phi_2 = \sigma - \varpi_2$.

The equations written above result from expanding the disturbing
function in eccentricities and averaging over time, so that only the
terms which do not vary rapidly with time are retained.  Therefore,
these equations will be valid as long as
$\phi_1$ and $\phi_2$ librate around some fixed values.  In the
numerical calculations we carry out in this paper, and starting from
initial conditions such that the system is close to MMR, we find that these
angles still librate around fixed values even 
when departure from exact MMR is significant.   This is probably due
to the fact that the orientation of the orbits becomes ``frozen'' when the
system evolves away from exact MMR.  As the interaction between the
planets weakens when they move away from MMR, there is no mechanism by
which these angles can be changed.  

\subsection{Modelling of migration and eccentricity damping}
\label{sec:migration}

Planets embedded in a disc are subject to both semimajor axis and
eccentricity damping on characteristic timescales $\tai$ and $\tei$,
respectively, where ${\rm i}=1,2$ refers to planets~1 and~2.  Note
that the {\em migration} timescale $\tmi$, over which the angular
momentum of the planets is damped, is such that $\tai=\tmi/2$ (e.g.,
Teyssandier \& Terquem 2014).  Damping of the semimajor axis
contributes an extra term $-\ai/\tai$ in the expression for
$\dot{a}_{\rm i}$, which is equivalent to adding $3 n_{\rm i} / ( 2 \tai)$ in the
expression for $\dot{n}_{\rm i}$.  Eccentricity damping is taken into account
by adding a damping term $-\ei / \tei$ in the expression for
$\dot{e}_{\rm i}$.  Eccentricity damping does in turn contribute to the
damping of the semimajor axis by a term $-2 \ai \ei^2 / \tei$, which
is equivalent to adding $3 n_{\rm i} \ei^2 / \tei$ in the expression for
$\dot{n}_{\rm i} $ (e.g., Teyssandier \& Terquem 2014).

With migration and eccentricity damping taken into account,
equations~(\ref{dn1dt})--(\ref{de2dt}) become:

\begin{align}
  \dot{n}_1 & = -3q n_1^2 \frac{\alpha m_2}{\ms} \left( e_1 f_1 \sin
                     \phi_1  + e_2 f^{\prime}_2 \sin \phi_2 \right) + \frac{3 n_1}{2
                     t_{a,1}} + \frac{3 n_1 e_1^2}{t_{e,1}},  \label{dn1dtm} \\
  \dot{n}_2 & = 3(q+1) n_2^2 \frac{m_1}{\ms} \left( e_1 f_1 \sin
                     \phi_1  + e_2 f^{\prime}_2 \sin \phi_2  \right) + \frac{3 n_2}{2
                     t_{a,2}} + \frac{3 n_2 e_2^2}{t_{e,2}}, \label{dn2dtm} \\
  \dot{e}_1 & = -n_1 \frac{\alpha m_2}{\ms}  f_1 \sin
                    \phi_1 - \frac{e_1}{t_{e,1}}, \label{de1dtm} \\
  \dot{e}_2 & = -n_2 \frac{m_1}{\ms}  f^{\prime}_2 
                    \sin \phi_2 - \frac{e_2}{t_{e,2}}. \label{de2dtm} 
\end{align}

In the regime of inward type--I migration that we focus on here, the
semimajor axis and eccentricity damping timescales can be written as:

\begin{equation}
  t_{a,{\rm i}}  = 27.0  \; \left[ 1+ \left( \frac{ e_{\rm i} }{1.3 H/r}\right)^5 
\right]
  \left[ 1- \left( \frac{ e_{\rm i} }{1.1 H/r}\right)^4 \right]^{-1} 
  \left( \frac{H/r}{0.02}
  \right)^2 \;
  \frac{{\rm M}_\odot}{m_{\rm d}} \;
  \frac{{\rm M}_\oplus}{m_{\rm i}} \; \frac{a_{\rm i}}{{\rm 1~au}}
  \left(  \frac{{\rm M}_\odot} {\ms} \right)^{1/2}   \;  {\rm yr} ,
\label{tm}
\end{equation}

and

\begin{equation}
t_{e,{\rm i}} =  3 \times 10^{-2} \left[ 1+ 0.25 \left( \frac{ e_{\rm i} }{H/r}\right)^3
  \right] \;  \left( \frac{H/r}{0.02}
  \right)^4  \\
  \frac{{\rm M}_\odot}{m_{\rm d}} \; \frac{{\rm M}_\oplus}{m_{\rm i}} \;
  \frac{a_{\rm i}}{{\rm 1~au}}  
\left( \frac{{\rm M}_\odot}{\ms} \right)^{1/2} \;  {\rm yr} ,
\label{te}
\end{equation}

\noindent (equations~[31] and~[32] of Papaloizou \& Larwood~2000, with
$f_s=1$ and $t_{a,{\rm i}} =t_m/2$).  Here $H/r$ is the disk aspect
ratio and $m_{\rm d}$ is the disk mass contained within 5~au.  The
equations assume that the disc surface mass density
$\propto r^{-3/2}$.

\section{Evolution of the system close to a resonance}
\label{sec:evolution}

Because the eccentricity damping timescales (eq.~[\ref{te}]) are much
smaller than the semimajor axis damping timescales (eq.~[\ref{tm}]),
the eccentricities quickly reach their equilibrium values.    We
calculate those values below, and then give expressions for the
evolution of the semimajor axes and departure from exact MMR.  

\subsection{Equilibrium values of the eccentricities}

\subsubsection{Convergent migration:}
\label{sec:econvergent}

We consider planets close to MMR and undergoing convergent migration
($t_{a,2} \le t_{a,1}$).   We assume that 
capture into the resonance is permanent,
%, which typically requires
%$m_2/\ms$ to be larger than $(t_{e,2}/t_{a,2})^{3/2} q/(q+1)^{3/2}$ 
 that is to say the damping timescales satisfy either
  equations~(26) or~(27) of Goldreich \&
Schlichting~(2014).  When the timescales do not statisty these
conditions, librations are overstable and the system escapes the
resonance on an eccentricity damping timescale. 

As can be seen from equations~(\ref{de1dtm}) and~(\ref{de2dtm}), the
eccentricities are being excited by the resonant interaction between the
planets and damped by the interaction of the planets with the disc.
When the planets are in MMR, $\dot{n}_1/n_1$ given by
equation~(\ref{dn1dtm}) is equal to $\dot{n}_2/n_2$ given by
equation~(\ref{dn2dtm}), which yields:

\begin{equation}
3qn_1 \left( e_1 f_1 \sin \phi_1 + e_2 f^{\prime}_2 \sin \phi_2 \right) \left(
  \frac{m_1}{\ms} + \frac{\alpha m_2 }{\ms} \right) = \frac{3}{2
  t_{a,1}} + \frac{3e_1^2}{ t_{e,1}} - \frac{3}{2
  t_{a,2}} - \frac{3e_2^2}{ t_{e,2}} . \label{eqecc}
\end{equation}

\noindent As we are looking for equilibrium values of the
eccentricities, we neglect the time derivatives in
equations~(\ref{de1dtm}) and~(\ref{de2dtm}), which yield:

\begin{align}
\frac{e_1}{t_{e,1}} & = -n_1 \frac{\alpha m_2}{\ms}  f_1 \sin \phi_1
                      , \label{sin1} \\
\frac{e_2}{t_{e,2}} & = -n_2 \frac{m_1}{\ms}   f^{\prime}_2 \sin \phi_2. \label{sin2}
\end{align}

\noindent Substituting into equation~(\ref{eqecc}) to
eliminate $\sin \phi_1$ and $\sin \phi_2$, we obtain:

\begin{equation}
  (q+1) \left(   1 +\frac{q}{q+1} \frac{m_1}{\alpha m_2} \right) \left( \frac{e_1^2}{ t_{e,1}} 
    + \frac{\alpha m_2}{m_1} 
      \frac{e_2^2}{t_{e,2}} \right) = -\frac{1}{2 t_{a,1}} + \frac{1}{2
        t_{a,2}} . \label{ecceq}
\end{equation}

\noindent When the planets are in MMR with no eccentricity damping,
the resonant angles librate around fixed values such that
$\sin \phi_{\rm i}=0$, so that $\phi_{\rm i}$ is equal to either
0$^{\circ}$ or 180$^{\circ}$ (eq.~[\ref{sin1}] and~[\ref{sin2}] with
$t_{e,1}$ and $t_{e,2}$ infinitely large).  With eccentricity damping,
these values are shifted by an amount which depends on the damping
timescales and mass ratios.  Equations~(\ref{sin1}) and~(\ref{sin2})
indeed yield
$\phi_{\rm i} \simeq e_{\rm i}/(t_{e,{\rm i}} m_{\rm i}/ \ms)$, where
$t_{e,{\rm i}}$ is in units of the orbital timescale and we have
replace $\sin \phi_{\rm i}$ by $\phi_{\rm i} $, assuming small angles.
Since at equilibrium the eccentricities are of order $(t_e/t_a)^{1/2}$
(see below), where $t_e$ and $t_a$ are typical damping timescales, we
obtain $\phi_{\rm i} \sim 1/ [ (t_et_a)^{1/2} m_{\rm i}/ \ms]$, where
here again the timescales are in units of typical orbital timescales.
Although in principle values of $\phi_{\rm i}$ may become large for
small mass ratios, we have checked that in the cases we investigate in
this paper it is reasonable to approximate these angles by 0$^{\circ}$
or 180$^{\circ}$.

We therefore assume that the resonant angles are close to some
fixed values such that any residual librations about these values may
be averaged out, i.e. $\dot{\phi}_1=\dot{\phi}_2=0$.  This implies
$\dot{\omega}_1=\dot{\omega}_2$ (and this is equal to
$\dot{\sigma}$).  Dividing equation~(\ref{dvarpi1dt}) by
equation~(\ref{dvarpi2dt}) then  yields:

\begin{equation}
\frac{e_2^2}{e_1^2} = \left(  \frac{m_1}{\alpha m_2} \right)^2  \left(
  \frac{q}{q+1} \right)^2  \left(  \frac{f^{\prime}_2}{f_1}
\right)^2, \label{eccratio}
\end{equation}

\noindent where we have replaced $\cos^2 \phi_1$ and $\cos^2 \phi_2$
by 1, as for the first--order resonances considered here the resonant
angles usually librate around 0$^{\circ}$ or 180$^{\circ}$.

\noindent Substituting into equation~(\ref{ecceq}), we then obtain
 the equilibrium value of $e_1^2$:

\begin{equation}
e_{1, {\rm eq}}^2 = 
\frac{t_{e,1}/t_{a,2}-t_{e,1}/t_{a,1}}{2 (q+1) \left( 1+ \frac{q}{q+1}
    \frac{ m_1}{\alpha m_2} \right)
\left[   1 +\frac{m_1}{\alpha
    m_2} \left( \frac{q}{q+1} \right)^2 \left(
    \frac{f^{\prime}_2}{f_1}  \right)^2 \frac{t_{e,1}}{t_{e,2}}
\right]}.
\label{e12eq}
\end{equation}

\noindent Note that this quantity is positive as $t_{a,1} > t_{a,2} $
(convergent migration).  

\noindent This formula is useful for calculating the equilibrium
values of the eccentricities, which are reached on a timescale much
shorter than the migration timescale, in any first order MMR.
Numerical values of $f_1$ and $f_2$ (and hence $f^{\prime}_2$) are given
in Appendix~\ref{app:coef} for $q$ between 1 and 6.

\noindent If $m_1 /m_2 \ll 1$, then using
$t_{a,1}/ t_{a,2} \sim t_{e,1}/ t_{e,2} \sim m_2/m_1$
(equations~[\ref{tm}] and~[\ref{te}]), equation~(\ref{e12eq}) can be
approximated by:

\begin{equation}
e_{1, {\rm eq}}^2 \simeq \frac{ t_{e,1}}{ 2(q+1) \left[  1+\frac{1}{\alpha}  \left( \frac{q}{q+1} \right)^2 \left(
    \frac{f^{\prime}_2}{f_1}  \right)^2 \right] t_{a,2} } \sim \frac {t_{e,1}}{t_{a,2} }.
\end{equation}

\subsubsection{Divergent migration:}
\label{sec:edivergent}

When the planets move away from MMR, the eccentricities are no longer
excited by the resonance and are only damped by the disc.
On a timescale on the order of $t_{e}$, they reach
values much smaller than the equilibrium value $\left( t_{e}/t_a
\right)^{1/2}$ found above (here $t_e$ and $t_a$ are typical
eccentricity and semimajor axis damping timescales), and ultimately
decay to zero. 

\subsection{Evolution of the mean motions}

\subsubsection{Convergent migration:}
\label{sec:mconvergent}

Using $(q+1)n_2=qn_1$, equations~(\ref{dn1dtm}) and~(\ref{dn2dtm}) can
be combined to give:

\begin{equation}
  \left( \frac{m_1}{\alpha m_2} +1 \right) \frac{\dot{n}_2}{n_2}  
  = \frac{ m_1}{ \alpha m_2} \frac{3}{2 t_{a,1}} +  \frac{3 }{2
    t_{a,2}} + \frac{3m_1}{ \alpha m_2} \left( \frac{ e_1^2}{t_{e,1}}
    + \frac{\alpha m_2}{m_1}  \frac{ 
    e_2^2}{t_{e,2}} \right).
\label{dn1dn2}
\end{equation}

\noindent 
Here again, we look for solutions with evolutionary long timescale,
large compared to the eccentricity damping timescales $t_{e,1}$ and
$t_{e,2}$, but smaller than the migration timescales $t_{a,1}$ and
$t_{a,2}$.  In this regime, the eccentricities are given by their
equilibrium values.  Substituting equation~(\ref{ecceq})
into equation~(\ref{dn1dn2}) then yields:

\begin{equation}
\frac{\dot{n}_2}{n_2}  = \frac{1}{t_n}, \label{dnidt2}
\end{equation}

\noindent with:

\begin{equation}
t_n = \frac{2 t_{a,2}}{3} \left(1 + \frac{q}{q+1} \frac{m_1}{\alpha
    m_2}\right)  \bigg/ \left( 1+\frac{q}{q+1}  \frac{m_1}{ \alpha
  m_2} \frac{t_{a,2}}{  t_{a,1} } \right).
\end{equation}

\noindent From equation~(\ref{tm}),  $t_{a,2}/t_{a,1} \propto a_2/a_1
$, which is constant, and $t_n$ depends on time through $t_{a,2}
\propto a_2$.  We write $t_n=t_{n,0} a_2/a_{2,0}$, where $a_{2,0}$ is
the initial value of $a_2$.  Using $\dot{n}_2/n_2=-3 \dot{a}_2/(2
a_2)$, the solution of equation~(\ref{dnidt2}) is then:

\begin{equation}
a_2=a_{2,0} \left( 1 - \frac{2t}{3t_{n,0}} \right),
\label{a2theory}
\end{equation}

\noindent and $a_1$ can be calculated using:

\begin{equation}
 a_1=a_2 \left( 1+\frac{1}{q} \right)^{-2/3}.
\label{a1theory}
\end{equation}

%\noindent When $m_1/m_2 \ll 1$, $t_n \simeq 2 t_{a,2}/3$.  In that case,
%$t_{a,1}$ is very large, as the migration timescale is inversely
%proportional to the planet mass for low mass planets, and the semimajor
%axis of both planets decreases over a timescale $t_{a,2}$.  

\subsubsection{Divergent migration:}
\label{sec:mdivergent}

As pointed out above, when the planets move away from MMR, the
eccentricities eventually get significantly smaller than their
equilibrium values.  The terms involving the eccentricities in
equations~(\ref{dn1dtm}) and~(\ref{dn2dtm}) then become negligible, as
does the gravitational interaction between the planets, and these
equations can be approximated by
$ \dot{n}_{\rm i} /n_{\rm i}=3/(2 t_{a,{\rm i}})$, with ${\rm i}=1,2$,
or, equivalently, $\dot{a}_{\rm i} /a_{\rm i}=-1/t_{a,{\rm i}}$.
Given that $t_{a,{\rm i}} \propto a_{\rm i}$, the solutions are:

\begin{equation}
a_{\rm i}  =  a_{{\rm i},0} \left( 1 - \frac{a_{\rm i}}{a_{{\rm i},0}}
  \frac{t}{ t_{a, {\rm i}}} \right),
\label{aidiv}
\end{equation}

\noindent where $a_{{\rm i},0}$ is the initial value of $a_{\rm i}$.

\subsection{Departure from exact MMR}
\label{sec:ratio}

\begin{figure}
\begin{center}
\includegraphics[scale=0.4]{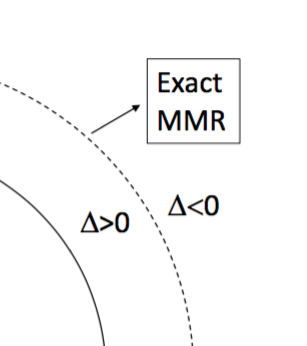}
\end{center}
\caption{Sketch illustrating the meaning of $\Delta$: the solid line
  shows the position of the inner planet, and the dashed line the
  position of the outer planet when the system is at exact MMR,
  corresponding to $\Delta=0$.  If the outer planet is interior
  (exterior) to the dashed line, $\Delta$ is positive (negative). }
\label{fig4}
\end{figure}

In this section, we no longer assume $n_2/n_1=q/(q+1)$, as was done in
the previous sections.  We define the parameter:

\begin{equation}
\Delta=(q+1) \frac{n_2}{n_1} -q,
\end{equation}

\noindent which measures the deviation from resonance.  If
$\Delta < 0$, $n_2/n_1 < q/(q+1)$, which means that the separation
between the planets is larger than at exact MMR (this is illustrated
in figure~\ref{fig4}).  Since the equations described in
section~\ref{sec:theory} and that we are using here are only valid
close to exact resonance, there is the implicit assumption that
$| \Delta | \ll 1$.

\noindent Using equations~(\ref{dn1dtm}) and~(\ref{dn2dtm}), we have:

\begin{align}
\frac{d \Delta}{dt} = & 3 (q+1) \frac{n_2}{n_1} \left( e_1 f_1 \sin \phi_1 +
  e_2 f^{\prime}_2 \sin \phi_2 \right)
  \left[ (q+1) n_2 \frac{m_1}{\ms} + q n_1 \frac{\alpha  m_2}{\ms}
  \right] \nonumber \\
& + 3 (q+1) \frac{n_2}{n_1} \left( \frac{1}{2t_{a,2}} +
  \frac{e_2^2}{t_{e,2}} - \frac{1}{2t_{a,1}} -
  \frac{e_1^2}{t_{e,1}} \right). \label{dotDelta1}
\end{align}

\noindent Once again, we look for solutions with evolutionary long
timescale, large compared to $t_{e,1}$ and $t_{e,2}$.  We therefore
can use equations~(\ref{sin1}) and~(\ref{sin2}) and combine them to
express
$\left( e_1 f_1 \sin \phi_1 + e_2 f^{\prime}_2 \sin \phi_2 \right)$ in
terms of the eccentricities.  Equation~(\ref{dotDelta1}) then becomes:

\begin{align} 
\frac{d \Delta}{dt} = & -3 (q+1) \frac{n_2}{n_1} \left[  (q+1)
  \frac{e_1^2}{t_{e,1}} \left( \frac{n_2}{n_1} \frac{m_1}{\alpha m_2}
    +1 \right) + q
  \frac{e_2^2}{t_{e,2}} \left( \frac{n_1}{n_2} \frac{\alpha m_2}{ m_1}
    +1 \right) \right] \nonumber \\
& +  \frac{n_2}{n_1} \frac{3(q+1)}{2} \left(
  \frac{1}{t_{a,2}} - \frac{1}{t_{a,1}} \right). \label{dotDelta2}
\end{align}

\noindent Here we assume that, although the planets may be moving away
from MMR, the resonant angles are still librating around some fixed
values, so that $\dot{\phi}_1=\dot{\phi}_2=0$.  We have checked that
this is indeed the case in all the numerical calculations we present
below, even when departure from exact MMR is significant.  This
implies $\dot{\omega}_1=\dot{\omega}_2=\dot{\sigma}$.  Using
equations~(\ref{dvarpi1dt}) and~(\ref{dvarpi2dt}) and
$\dot{\sigma}=(q+1)n_2 - qn_1 = n_1 \Delta$, this yields:

\begin{align}
e_1^2 = & \left( \frac{\alpha m_2}{\ms} \right)^2
          \frac{f_1^2}{\Delta^2},  \label{e12} \\
e_2^2 = & \left( \frac{ m_1}{\ms} \right)^2 \left( \frac{n_2}{n_1} \right)^2
          \frac{ f^{\prime 2}_2}{\Delta^2}, \label{e22}
\end{align}

\noindent where we have replaced $\cos^2 \phi_1$ and $\cos^2 \phi_2$
by 1.  Substituting into equation~(\ref{dotDelta2}) and using
$n_2/n_1 = (\Delta +q)/(q+1)$, we finally obtain the following
differential equation:

\begin{equation}
\Delta^2 \frac{d \Delta}{dt} = -{\cal{A}} -{\cal{B}} \Delta^2, \label{dotDelta3}
\end{equation}

\noindent with:

\begin{align}
  {\cal{A}} = &  \frac{3(\Delta +q)}{t_{e,1}} \left( \frac{\Delta
                          +q}{q+1} \frac{m_1}{\alpha m_2} +1 \right)  \left( \frac{ \alpha
                          m_2}{\ms} \right)^2 
%                          \nonumber \\ & \times 
                                          \left[  (q+1) f_1^2
                                          +
                                          \frac{q(\Delta +q)}{q+1} \frac{m_1}{\alpha m_2}
                                          f^{\prime 2}_2                            
                                          \frac{t_{e,1}}{t_{e,2}}
                                          \right], \label{coefA} \\ 
  {\cal{B}} =            &  \frac{3(\Delta
                           +q)}{2t_{a,1}} \left( 1 -
                           \frac{t_{a,1}}{t_{a,2}}  \right).  \label{coefB}
\end{align}

\noindent As $\Delta+q=(q+1)n_2/n_1 >0$, ${\cal{A}}$ is a positive
definite function of $\Delta$. 

%\noindent Setting the new variable $z \equiv \Delta^3$,
%equation~(\ref{dotDelta3}) can be written as:

%\begin{equation}
%\frac{dz}{dt}= - 3 {\cal{A}} - 3 {\cal{B}} |z|^{2/3}.
%\label{dotz}
%\end{equation}

%\noindent Since $n_2 < n_1$, we have $-q < \Delta <1$, and therefore
%$-q^3 < z < 1$.
 
\noindent We now discuss solutions of this equation in the case of
both convergent and divergent migration, adopting a model where
$t_{a, {\rm i}}$ and $t_{e, {\rm i}}$ are proportional to the
semimajor axis (eq.~[\ref{tm}] and~[\ref{te}]), which results from
assuming that the disc surface mass density $\propto r^{-3/2}$.

\subsubsection{Convergent migration:}
\label{sec:convergent}

In this regime, $t_{a,2} \le t_{a,1}$, and ${\cal{B}}$ given
by equation~(\ref{coefB}) is negative.  Therefore
equation~(\ref{dotDelta3}) has equilibrium solutions with $\Delta$
constant such that
$\Delta^2 = -{\cal{A}}/{\cal{B}} $, where ${\cal{A}}$ and
${\cal{B}}$ are functions of $\Delta$. 

\noindent Assuming small departure from resonance (which can be
justified {\em a posteriori}), which implies $|\Delta| \ll q$ and
therefore ${\cal{A}}$ and ${\cal{B}}$ independent of $\Delta$,
${\cal{A}}/{\cal{B}}$ can be expressed in terms of the masses and
damping timescales.  For convergent migration to occur, $m_1/m_2$ has
to be on the order of unity or smaller.  In that case, using
$t_{a,1}/t_{e,1} \sim 10^3$ (eq.~[\ref{tm}] and~[\ref{te}]),
$ \sqrt{ - {\cal{A}}/{\cal{B}}  }$ can be evaluated from
equations~(\ref{coefA}) and~(\ref{coefB}) and is found to be very
small compared to unity for Earth mass planets.  Therefore, to a high
degree of accuracy, the planets can be considered in exact
MMR. 

\noindent Equilibrium values of $\Delta$ can either be positive or
negative, equal to $\pm \sqrt{ -{\cal{A}}/{\cal{B}} }$.  If $\Delta$
starts from an initial value exceeding the positive root of smallest
magnitude, the right--hand side of equation~(\ref{dotDelta3}) is
positive, which implies that $d \Delta /dt >0$ and $\Delta$ moves away
from this root.  In the same way, $\Delta$ cannot converge toward the
positive root from below.  Convergence towards the negative root
though is possible, both from below and from above, and therefore this
is an  equilibrium solution which can be approached time
asymptotically.  This corresponds to the formation and maintenance of
a commensurability in which the planets are slightly {\em further
  apart} than at exact MMR.

\noindent If $\Delta$ is initially slightly larger than the positive
root of smallest magnitude, as we have just pointed out, it diverges
from it.  In principle, there is the possibility of larger positive
roots corresponding to equilibrium that could be approached time
asymptotically if the migration timescales had an appropriate
dependence on the orbital parameters.  This is illustrated below in
section~\ref{sec:numconv}.

\subsubsection{Divergent migration: an approximate solution}
\label{sec:divergent}

We now consider the case where migration is divergent, which happens
when $t_{a,2} > t_{a,1}$.  

\noindent In that case, ${\cal{B}}$ given by equation~(\ref{coefB}) is
a positive definite function of $\Delta$, and $\Delta$ given by
equation~(\ref{dotDelta3}) must decrease with time, corresponding to
an expansion away from commensurability.  There is no steady state in
this case.  The first term on the right--hand--side of
equation~(\ref{dotDelta3}) corresponds to expansion due to orbital
circularization, and it dominates for small $\Delta$.  The second term
corresponds to divergent migration and it dominates for larger values
of $\Delta$. 

\noindent The solution of equation~(\ref{dotDelta3}) is:

\begin{equation}
\Delta - \int_0^{\Delta} \frac{ ( {\cal{A}} / {\cal{B}} ) d \Delta'}{ (
  {\cal{A}} / {\cal{B}}  ) +
  \Delta'^2  } = - \int_0^t
{\cal{B}}(t') dt' ,
\label{deltap1}
\end{equation}

\noindent where we have assumed that $\Delta=0$ (exact
commensurability) at $t=0$.  Provided
$ \left| {\cal{A}} / {\cal{B}} \right| \ll 1 $, most of the
contribution to the integral on the left--hand--side comes from near
$\Delta'=0$.  Therefore, when calculating this integral, we approximate
${\cal{A}}$, ${\cal{B}}$ and other orbital parameters by their value
at the center of the resonance (i.e. at $\Delta=0$).  Under these
circumstances, ${\cal{A}}$ and ${\cal{B}}$ depend on orbital
parameters through ${\cal{A}} \propto 1/t_{e,1} \propto 1/a_1$ and
${\cal{B}} \propto 1/t_{a,1} \propto 1/a_1$.

Equation~(\ref{deltap1}) can then be written as:

\begin{equation}
\Delta - \sqrt{\frac{ {\cal{A}} }{{\cal{B}}}} \tan^{-1} \left(
  \sqrt{\frac{{\cal{B}}}{{\cal{A}}}} \Delta   \right) = - \int_0^t
{\cal{B}}(t') dt'.
\label{deltap}
\end{equation}

\noindent For small $t$ and hence $\Delta$, ${\cal{B}}(t') $ on the
right--hand side of this equation may in addition be assumed to be
constant and equal to its value for $\Delta=0$ and
$a_1=a_{1,0}$. Furthermore, we expand the left--hand--side in
powers of $\Delta$, retaining only the first non vanishing term.  
This yields:

\begin{equation}
\Delta=-\left(   3 {\cal{A}} t  \right)^{1/3}.
\label{deltaA}
\end{equation}

\noindent This time--dependence of $t^{1/3}$ is in agreement with
previous results (Papaloizou \& Terquem ~2010, Lithwick \& Wu~2012 and
Batygin \& Morbidelli~2013).  

\noindent For larger values of $t$, ${\cal{B}}(t') $ on the
right--hand--side of equation~(\ref{deltap1}) can no longer be assumed
to be constant, whereas the $\tan^{-1}$ term becomes negligible
compared to $\Delta$.  We write ${\cal{B}}=({\cal{B}}a_1)/a_1$, with
the numerator being a constant.  As the planets are moving away from
MMR, $a_1$ is given by equation~(\ref{aidiv}) and we can write:

\begin{equation}
\Delta = - \int_0^t {\cal{B}}(t') dt' = - \int_0^t \frac{{\cal{B}}
  t_{a,1} \; dt'}{(a_{1,0}/a_1) t_{a,1}-t'}  = {\cal{B}} t_{a,1} \ln \left|
  1 - \frac{a_1}{a_{1,0}}\frac{t}{t_{a,1}} \right|.
\label{deltaB1}
\end{equation}

\noindent As this expression only becomes valid after some time that
we note $t_{\rm ref}$, and at which $\Delta=\Delta_{\rm ref}$, we
substract the form of equation~(\ref{deltaB1}) for $t=t_{\rm ref}$
from the form at time $t$, so that we finally obtain:

\begin{equation}
  \Delta={\cal{B}} t_{a,1}  \ln \left| \frac{ 
  1 - \frac{a_1}{a_{1,0}} \frac{t}{t_{a,1}}
 }{ 
  1 - \frac{a_1}{a_{1,0}} \frac{t_{\rm ref}}{t_{a,1}}
} \right|
+\Delta_{\rm ref}.
\label{deltaB}
\end{equation}

\section{Numerical solutions}
\label{sec:numeric}

We solve equations~(\ref{dvarpi1dt})--(\ref{de2dtm}) for the variables
$\varpi_1$, $\varpi_2$, $\sigma$, $n_1$, $n_2$, $e_1$, $e_2$, $\phi_1$
and $\phi_2$.
For illustrative purposes, we fix $q=2$, as the 3:2 MMR is common
among observed systems.

\subsection{Convergent migration}
\label{sec:numconv}

Convergent migration requires $t_{a,1} > t_{a,2}$ .  From
equation~(\ref{tm}), this implies that
$m_2/m_1 > a_2/a_1 = (1+1/q)^{2/3}=1.3$.  We choose $m_1= 1$~$\me$ and
$m_2=1.4$~$\me$, which satisfies this condition.  We adopt $H/r=0.02$,
$m_{\rm d}=10^{-4}$~$\msun$ and $\ms=1$~$\msun$.  We start the inner
planet at $a_1=2.5$~au and the outer planet at
$a_2=a_1(1+1/q)^{2/3}=3.26$~au.  The initial timescales are then, in
years, $t_{e,1}=7.5 \times 10^2$, $t_{a,1}=6.7 \times 10^5$,
$t_{e,2}=7 \times 10^2$ and $t_{a,2}=6.3 \times 10^5$.  With these
values, the criterion for permanent capture in resonance with finite
libration amplitude, which is
$m_2/ \ms > 0.17(t_e/t_a)^{3/2}q/(q+1)^{3/2} $, where $t_e$ and $t_a$
are typical eccentricity and semimajor axes damping timescales, is
satisfied (Goldreich \& Schlichting~2014).  The initial values of
$e_1$, $e_2$, $\varpi_1$, $\varpi_2$ and $\sigma$ are chosen
arbitrarily as this does not affect the calculations.

\begin{figure}
\begin{center}
\includegraphics[scale=0.7]{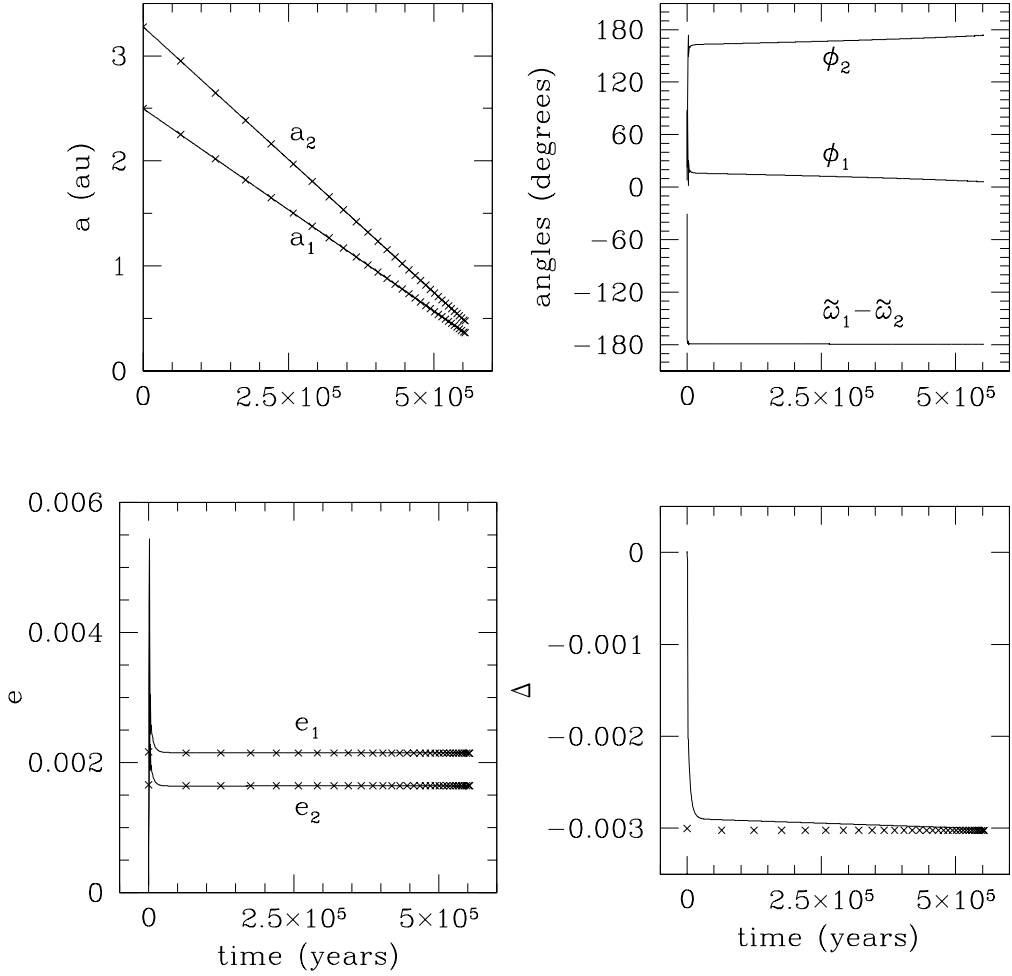}
\end{center}
\caption{Evolution of the system for $q=2$, $m_1= 1$~$\me$ and
  $m_2=1.4$~$\me$ (convergent migration) as a function of time (in years).  
The upper left--hand plot
  shows the semimajor axes (in au) corresponding to the
  numerical calculations (solid lines) and to the
  analytical results given by equations~(\ref{a2theory})
  and~(\ref{a1theory}) (crosses).
The upper
  right--hand plot shows the resonant angles $\phi_1$ and $\phi_2$ and
  the difference of the pericenter longitudes $\varpi_1-\varpi_2$ (in
  degrees). The lower left--hand plot
  shows the
  eccentricities corresponding to the
  numerical calculations (solid lines) and to the
  analytical results given by equations~(\ref{e12eq})
  and~(\ref{eccratio}) (crosses).  The lower right--hand plot shows
  $\Delta$ corresponding to the numerical calculations (solid line)
  and to the analytical result  $- \sqrt{ -{\cal{A}}/{\cal{B}} }$ with
${\cal{A}}$ and ${\cal{B}}$ given by equations~(\ref{coefA})
and~(\ref{coefB}) (crosses).}
\label{fig1}
\end{figure}

Figure~\ref{fig1} shows the numerical results and a comparison with
the analytical results.  The semimajor axes decrease while maintaining
commensurability, and their evolution is in excellent agreement with
that given by equations~(\ref{a2theory}) and~(\ref{a1theory}).  As
expected, the resonant angles are close to 0$^{\circ}$ or
180$^{\circ}$, whereas the difference of the pericenter longitudes
$\varpi_1-\varpi_2$ is close to $-180 ^{\circ}$.  Both eccentricities
reach equilibrium values which are also in excellent agreement with
equations~(\ref{e12eq}) and~(\ref{eccratio}).  The parameter $\Delta$
decreases slightly from the initial value of 0, and reaches the
equilibrium solution $- \sqrt{ -{\cal{A}}/{\cal{B}} }$, as predicted
in section~\ref{sec:convergent}.

\begin{figure}
\begin{center}
\includegraphics[scale=0.7]{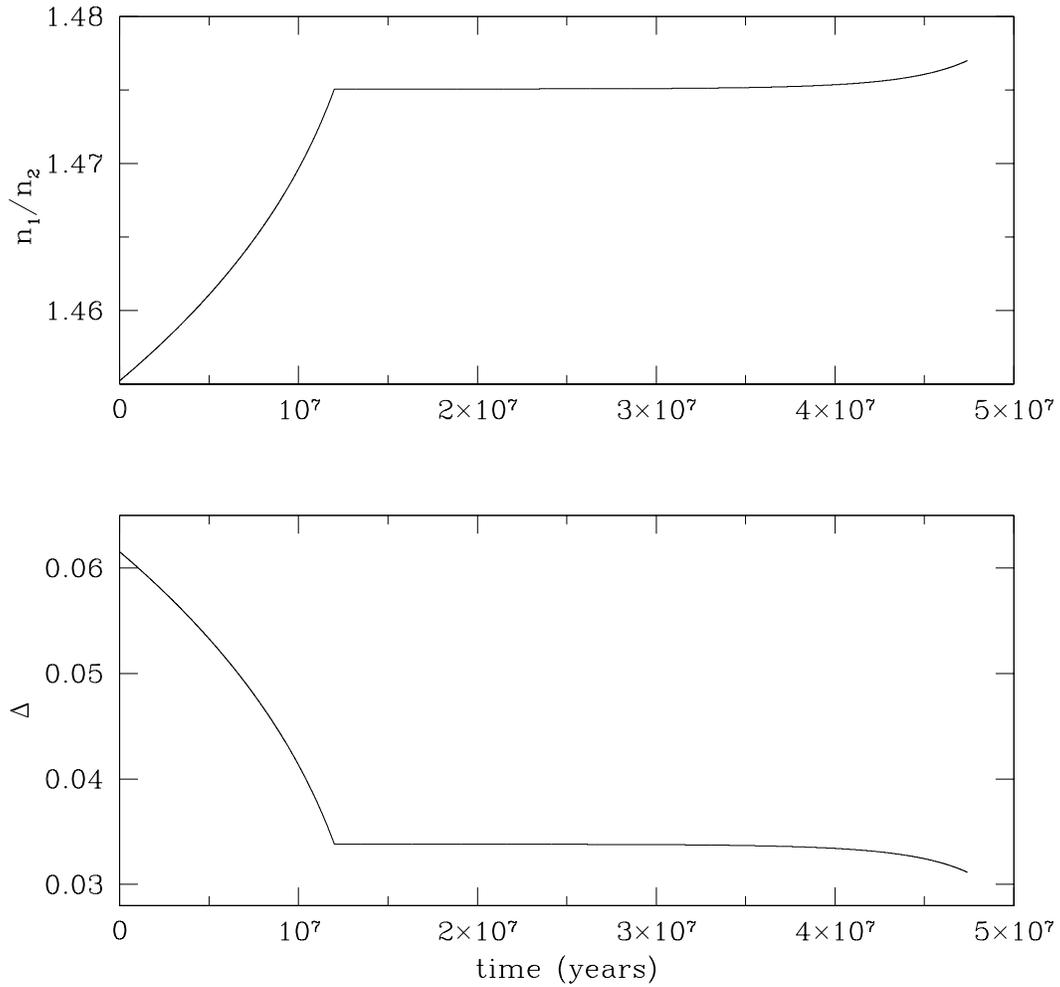}
\end{center}
\caption{Evolution of the system for the same parameters as in
  figure~\ref{fig1}, except that here the timescales are adjusted in
  such a way that the planets stall {\em interior} to exact MMR.  The
  plots show the ratio $n_1/n_2$ (upper plot) and $\Delta$ (lower
  plot) {\em versus} time (in years).  The planets start slightly
  interior to the 3:2 MMR.  The system is then evolved with no
  migration but eccentricity damping timescales
  $t_{e,1}=t_{e,2} \simeq 4 \times 10^2$~years, so that the evolution
  is towards MMR.  At $t=1.2 \times 10^7$~years, before exact MMR is
  reached, a convergent migration timescale is applied which cancels
  out the expansion of the system.  This results in the planets
  stalling interior to exact MMR for several $10^7$~years. }
\label{fig2}
\end{figure}

We have pointed out above that equilibrium solutions with $\Delta >0$,
which correspond to the planets being {\em interior} to the MMR, could
be attained if the migration timescales were adjusted appropriately.
This is illustrated in figure~\ref{fig2}.  In this calculation, the
planets start slightly interior to the 3:2 MMR, and the system is
evolved with no migration but eccentricity damping timescales
$t_{e,1}=t_{e,2} \simeq 4 \times 10^2$~years.  This produces an
increase of the mean motions, with $n_1$ evolving faster than $n_2$
(see eq.~[\ref{dn1dtm}] and~[\ref{dn2dtm}]), so that the system
evolves towards MMR.  If the system kept evolving under the action of
eccentricity damping only, it would pass through MMR and keep
separating.  However, before exact MMR is reached, we apply a
convergent migration timescale about 100 times longer than that given
by equation~(\ref{tm}) which cancels out the expansion of the system,
so that the planets stall interior to exact MMR for several
$10^7$~years.  This timescale could be increased by refining the
migration timescale.  As shown in figure~\ref{fig2}, departure from
exact MMR is significant. However, we have checked that the resonant
angles still librate around 0$^{\circ}$ and 180$^{\circ}$ in that
case, so that equations~(\ref{dvarpi1dt})--(\ref{de2dtm}) can still be
used to describe the evolution of the system.  The situation we have
described here is not very realistic but illustrates that equilibrium
positions {\em interior} to MMR could be reached.  Such a case could
arise if, for example, movement towards MMR as a result of orbital
circularization were balanced by very weak convergent migration.

\subsection{Divergent  migration}

Here we adopt $m_1=1$~$\me$ and $m_2=1.3$~$\me$ together with $m_{\rm
  d}=5 \times 10^{-6}$~$\msun$, which ensures that the planets 
move away from MMR on a long timescale, and we keep $\ms=1$~$\msun$
and $H/r=0.02$.  The initial values of $a_1$ and $a_2$ are the same as
above.  Given that the planet masses are very close to the values they
had in the convergent case above, but the disc mass
is 20 times smaller, all the damping timescales are 20 times longer.

\begin{figure}
\begin{center}
\includegraphics[scale=0.7]{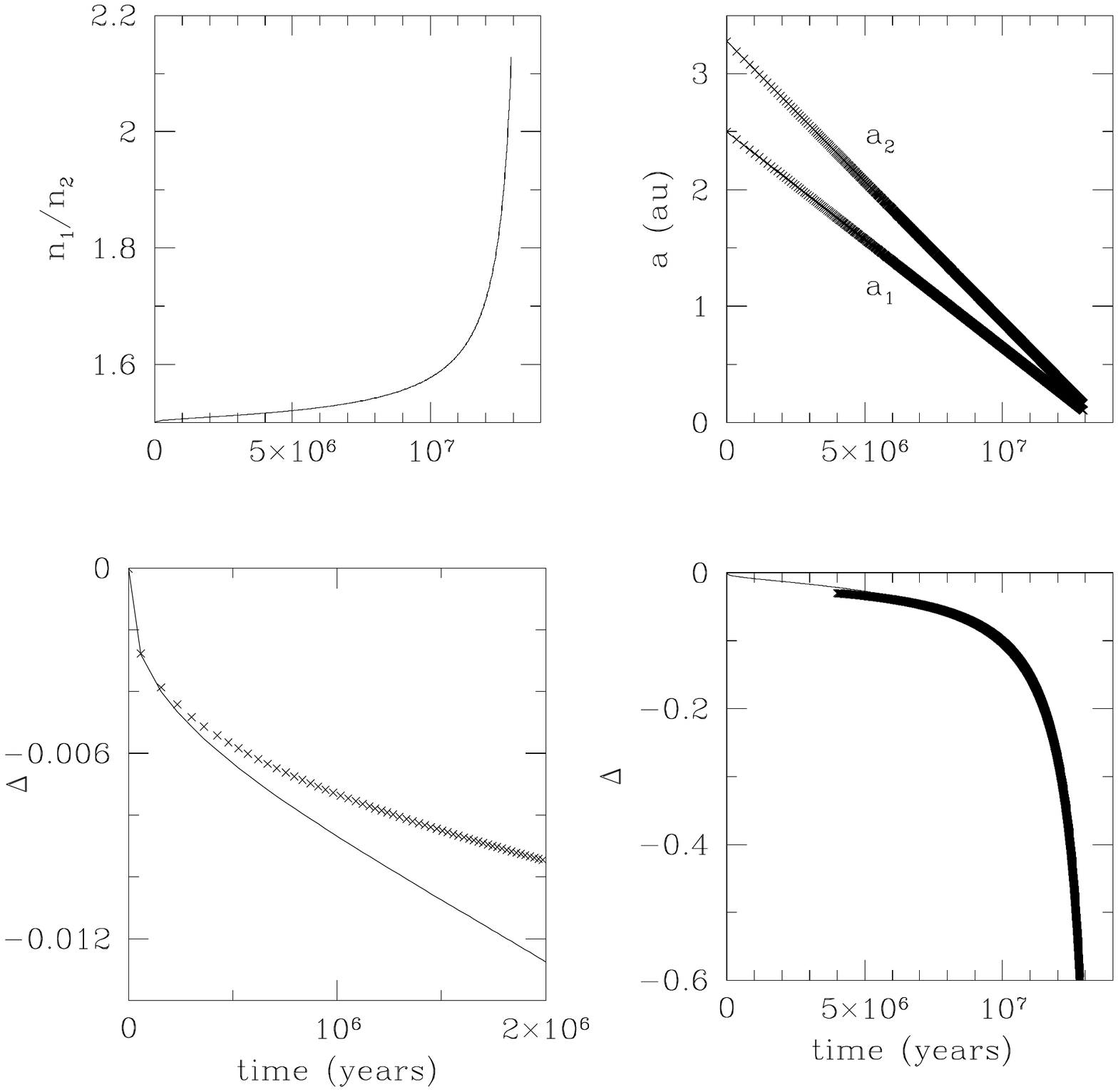}
\end{center}
\caption{ Evolution of the system for $q=2$, $m_1= 1$~$\me$ and
  $m_2=1.3$~$\me$ (divergent migration).  The upper left--hand plot
  shows $n_1/n_2$ {\em versus} time (in years).  The upper
  right--hand plot shows the semimajor axes (in au)
  {\em versus} time (in years). The solid lines correspond to the
  numerical calculations whereas the crosses correspond to the
  analytical results given by equation~(\ref{aidiv}). The lower plots show $\Delta$
  {\em versus} time (in years). The left--hand plot is a zoom on early
  times.  The solid lines correspond to the
  numerical calculations whereas the crosses correspond to the
  analytical results given by equations~(\ref{deltaA}) (left--hand plot)
  and~(\ref{deltaB}) (right--hand plot).}
\label{fig3}
\end{figure}

Figure~\ref{fig3} shows the evolution of $n_1/n_2$, of the semimajor
axes and of $\Delta$ for this system.  At the beginning of the
evolution, $\Delta$ decreases as $t^{1/3}$, in agreement with
equation~(\ref{deltaA}), whereas at later times it evolves as
predicted by equation~(\ref{deltaB}).

We have checked that the eccentricities rapidly decrease to values
very small compared to the equilibrium values, and that $\phi_1$ and
$\phi_2$ keep librating around fixed values, even for such large
departures from exact MMR, so that the analysis done in
sections~\ref{sec:mdivergent} and~\ref{sec:ratio} applies.

\section{Observed systems}
\label{sec:observations}

Using the {\rm Open Exoplanet Catalogue}\footnote{{\em
    http://openexoplanetcatalogue.com/}}, we have selected all the two
planet systems in which the planet radii are smaller than
3.92~$\rearth$ and the period ratio $P_2/P_1$ is smaller than 2.3.
The constraint on the radius ensures that the planets are likely to be
Earths or super--Earths and in the regime of type~I migration.  The
constraint on the period ratio selects for systems which are close to
MMR.  There are~107 such systems.  We have also included in our sample
9~systems with more than 2 planets but in which a pair of planets
satisfies the above criteria and is far enough from the other planets
that it is not expected to interact significantly with them.

\subsection{Departure from exact resonances}

The fact that $P_2/P_1 \le 2.3$ means that there exists an integer
$q \ge 1$ such that $-0.25 \le P_2/P_1 - (q+1)/q \le 0.3$.  Here we
focus on first--order MMRs as these are the resonances in which
low--mass planets are most easily captured (Papaloizou \&
Szuszkiewicz~2005).  The first--order MMR the system is closest to is
the (q+1):q resonance where $q$ is such that it minimizes
$\left| P_2/P_1 - (q+1)/q \right|$.  The departure from exact
resonance is then $\delta \equiv P_2/P_1 - (q+1)/q$.  However,
according to this criterion, which we label~(a), the system can either
be interior ($\delta < 0$) or exterior ($\delta > 0$) to the
resonance.  As it has been argued that systems can more easily be
produced exterior rather than interior to resonances, when
$\delta <0$, we increase $q$ by 1, which makes $\delta >0$, as long as
this change keeps $\delta <0.3$.  This criterion is labelled~(b).  For
example, $P_2/P_1=1.78$ corresponds to $q=1$ according to
criterion~(a), which gives $\delta=-0.22$.  If we use criterion~(b)
instead, we have $q=2$ and $\delta=0.28$.  Therefore the system can be
seen as being either interior to the 2:1 MMR or exterior to the 3:2
MMR.  The choice of 0.3 as a threshold for $\delta$ is completely
arbitrary, but it has been taken large enough to favour systems
exterior rather than interior to resonances, and small enough that the
departure from exact MMR is within 15\%.

\noindent Note that the relation between $\delta$ and $\Delta$ defined
above is:

\begin{equation}
\Delta=\frac{-q \delta}{\delta + (q+1)/q}.
\end{equation}

\begin{figure}
\begin{center}
\includegraphics[scale=0.7]{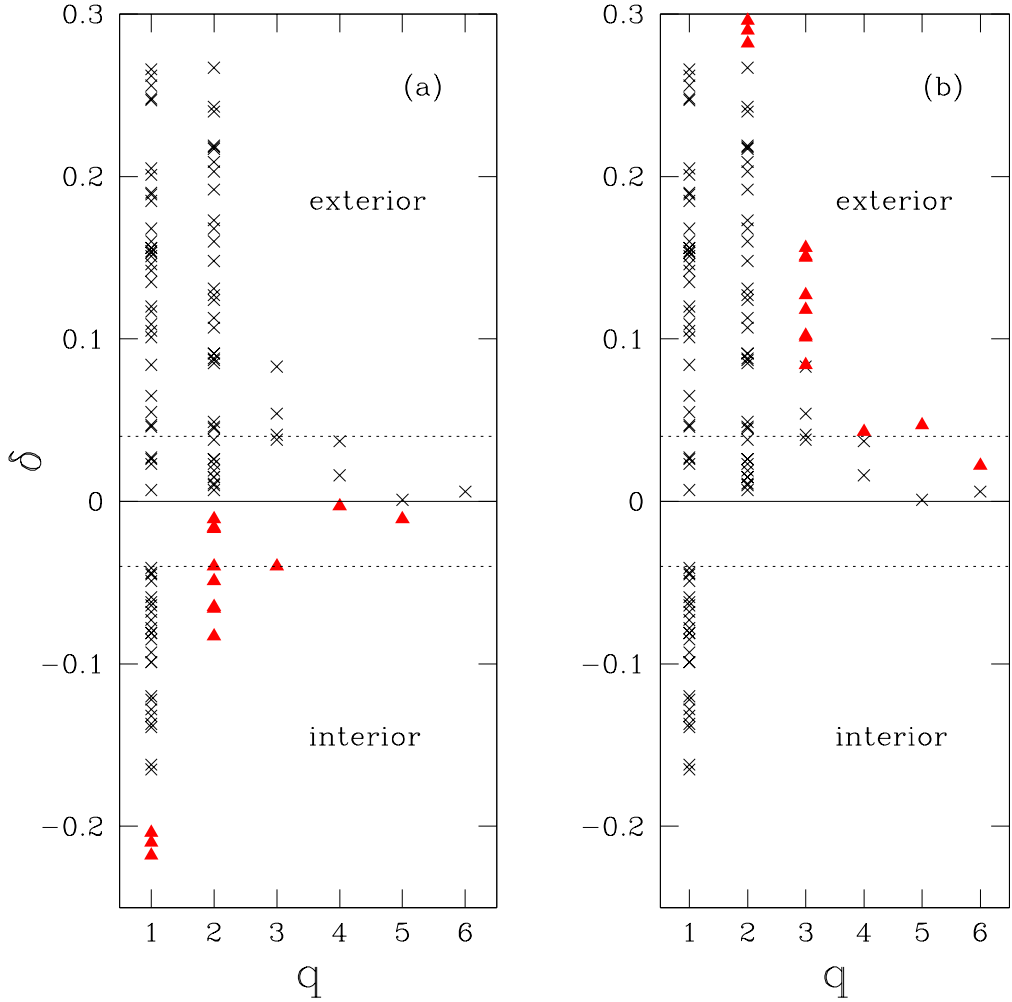}
\end{center}
\caption{ Departure from exact MMR, as measured by
  $\delta \equiv P_2/P_1-(q+1)/q$, {\em versus}
  $q$.  For the left--hand plot, which is labelled (a), the value of
  $q$ is chosen in such a way as to minimize $\left| \delta \right|$.
  For the right--hand plot, which is labelled (b), the value of $q$
  is chosen in such a way as to favour positive values of $\delta$
  while keeping $\left| \delta \right|<0.3$.  This amounts to
  increasing $q$ by 1 from the left to the right plots for some of the
  points, which are coloured in red.  Exact MMR corresponds to $\delta=0$,
  whereas $\delta < 0 $ ($\delta >0$) corresponds to planets interior (exterior)
  to the resonance. The dashed lines represent $\delta= \pm 0.04$. }
\label{fig5a}
\end{figure}

\noindent Figure~\ref{fig5a} shows $\delta$ as a function of $q$ for
all the systems in our sample, with $q$ chosen according to either
criterion~(a) or~(b). (Similar plots have been published by Steffen \&
Hwang~2015, but without considering separately two--planet systems).
Since the distance between MMR for $q \ge 2$ is less than
$3/2-4/3 \simeq 0.17$, all the systems with $P_2/P_1 \le 1.8$ can be
seen as exterior to a MMR with $q \ge 2$ according to criterion~(b).
However, this is not true for $q=1$, and there are systems interior to
the 2:1 resonance.  Using criterion~(a), there are 61 systems with
$q=1$, 27~of them with $\delta<0$ and 34~with $\delta >0$.  If we use
criterion~(b) instead, 3~of those systems with $\delta <0$ are
assigned $q=2$ rather than $q=1$.  Therefore, even if we adopt a
criterion that favours planets being exterior rather than interior to
MMR, we see that more than 40\% of the systems close to the 2:1
resonance are interior to it.
%As there is no reason to believe that this resonance is special
%compared to the other first--order MMR, it is likely that
%criterion~(a) is more reliable than criterion~(b), and pairs of
%planets are not preferentially exterior to MMR.

We see in figure~\ref{fig5a} that when planets are very close to a
resonance ($|\delta| < 4\% $), they tend to be exterior rather than
interior to the resonance, in agreement with Fabrycky et al. (2014).
However, for even slightly larger departures, the spread is in either
direction.

Labelling a system as exterior rather than interior to a resonance may
seem like a semantic issue.  However, it may be that the physical
processes that move a system in one or the other direction from exact
MMR are different.  For example, it has been pointed out that
dissipation of energy at constant angular momentum always moves the
system further apart (Papaloizou \& Terquem~2010, Papaloizou~2011,
Lithwick \& Wu~2012, Delisle et al. 2012, Batygin \& Morbidelli~2013),
so that it ends up being slightly exterior to the resonance.  It is
therefore of interest to try to understand whether the data indicate a
tendency or not.

\begin{figure}
\begin{center}
  \includegraphics[scale=0.5]{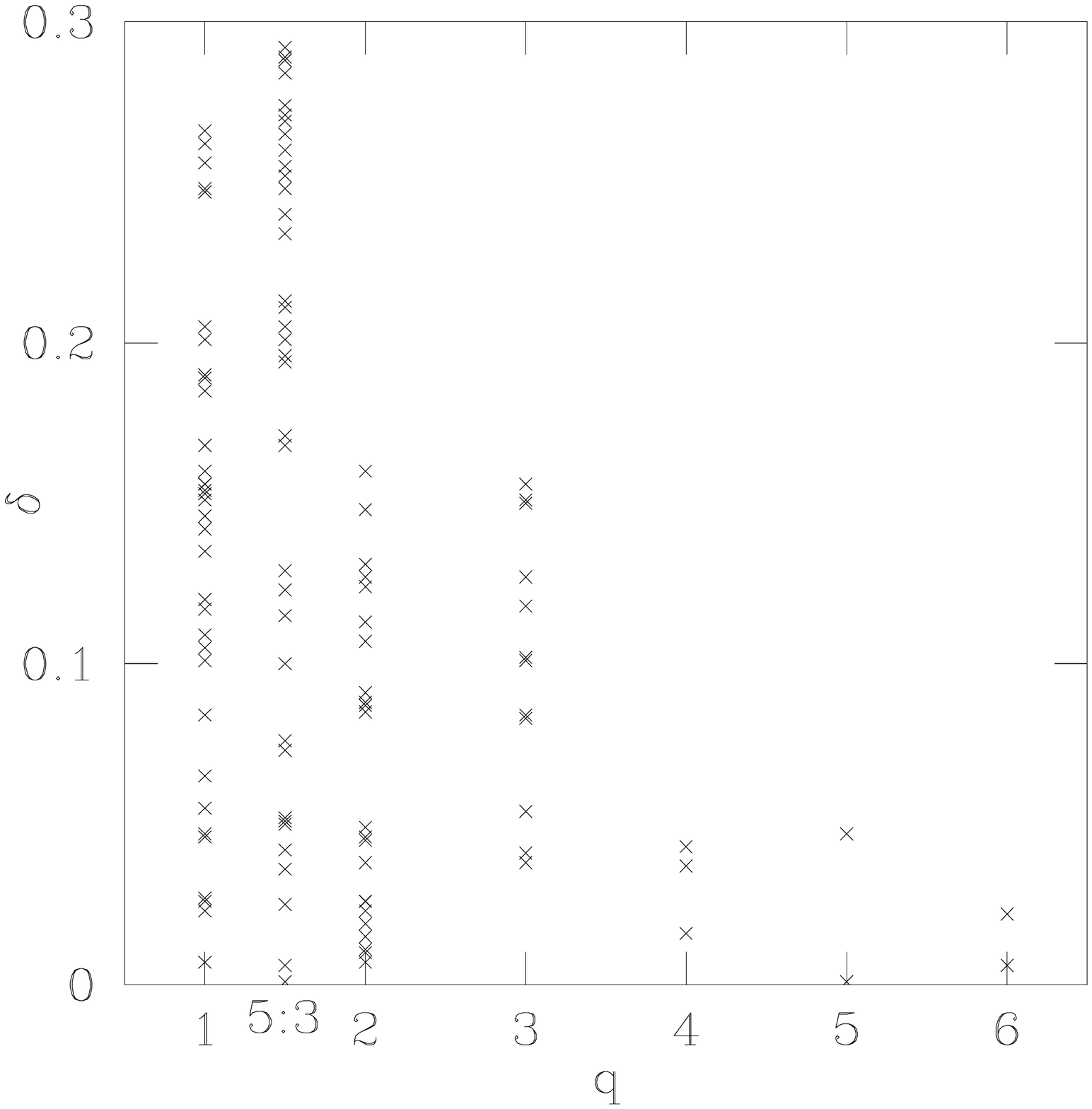}
\end{center}
\caption{ Same as the right--hand plot of figure~\ref{fig5a} but with
  the second--order 5:3 resonance now included.   
All the systems which had $q=1$ and $\delta<0$ are now
assigned  the 5:3 resonance, as are the systems which had $q=2$ and
$\delta>0.167$.}
\label{fig5b}
\end{figure}

Figure~\ref{fig5a} includes only first--order MMRs.  As the 5:3
second--order MMR is located 0.33 below the 2:1 MMR, all the systems
interior to the 2:1 resonance could be seen as being exterior to the
5:3 resonance.  In figure~\ref{fig5b}, we again show $\delta$ as a function
of $q$ but we now include the 5:3 MMR.  We adopt criterion~(b), which
means that all the systems which had $q=1$ and $\delta<0$ are now
assigned  the 5:3 resonance, as are the systems which had $q=2$ and
$\delta>0.167$.

\noindent Adding the 5:3 resonance enables to argue that all the
systems are exterior to a resonance.  However, this implies that the
number of systems captured in the 5:3 MMR is comparable to that
captured in the 3:2 MMR, which is not consistent with the result of
numerical simulations for low--mass planets (Xiang--Gruess \&
Papaloizou~2015).  Also, it forces us to consider that systems that
are very close but interior to the 2:1 MMR are in fact captured in the
5:3 resonance and moved rather far away outside that resonance.

As has been pointed out in previous studies (Lissauer et al. 2011,
Fabrycky et al. 2014), a striking feature of either figure~\ref{fig5a}
or figure~\ref{fig5b} is that there is a large spread of $\delta$
around each MMR, even if we add the 5:3 resonance.  From the figures
above, we can conclude that either:
\begin{itemize}
\item[(i)] all the systems attain exact MMR through smooth convergent
  migration; a few of them are subsequently moved exterior to the
  resonance by less than 4\% or so by some 'gentle' processes, while
  the majority of the systems are moved significantly away from the
  MMR in either direction by some other more efficient processes;
\item[(ii)] only a small fraction of the systems attain exact MMR
  and are moved exterior to the resonance by less than 4\% or so by
  some 'gentle' processes; the vast majority of the systems are
  distributed randomly and were never captured in resonances.
\end{itemize}
In both scenarii, systems that would have found themselves slightly
interior to a resonance by less than 4\%, i.e. with $-0.04<\delta<0$,
would be moved toward positive $\delta$ by the 'gentle' process.  

\noindent The 'gentle' processes we refer to may involve dissipation
of energy at constant angular momentum.  More efficient processes that
could take a system of two planets further away from resonance have
been proposed but it is not clear so far that any of them can explain
the range of data.  This is investigated in more details in
section~\ref{sec:cavity} and discussed in
section~\ref{sec:discussion}.

\subsection{Convergent {\em vs} divergent migration}

For most of the planets in our sample, a radius but not a mass has
been measured.  There is no unique relation between mass and radius
for Earth and super--Earth like planets, as they span a wide range of
compositions (Baraffe et al. 2014).  However, some probabilistic
mass--radius relations have been proposed based on a sample of well
constrained planets (Wolfgang et al.~2016, Chen \& Kipping~2017).  In
order to obtain a mass for the planets in our sample when there is no
value derived from observations, we use the relations proposed by Chen
\& Kipping~(2017):

\begin{align}
\frac{m_p}{\me} & =0.952 \left( \frac{r_p}{\rearth} \right)^{3.584}
                  \; \; \; {\rm for} \;
  r_p \le 1.23 \; \rearth, \\
\frac{m_p}{\me} & = 1.407 \left( \frac{r_p}{\rearth} \right)^{1.698} \; \; \; {\rm for} \;
  r_p \ge 1.23 \; \rearth, 
\end{align}

\noindent where $m_p$ and $r_p$ are the mass and the radius of the
planet.  Comparing $m_p$ obtained from these relations with the
observed value when it exists shows that these relations do not
give very good individual fits.  However, we note that in this paper
we are more interested in the ratio of the masses than in the masses
themselves, as it is the ratio that determines whether migration is
convergent or divergent.  Using either the observed mass when it
exists of that
derived using the relations above, together with equation~(\ref{tm}), we
calculate the ratio of the migration timescales that corresponds to the
planets being in exact MMR:

\begin{equation}
\frac{t_{a,2}}{t_{a,1}} = \frac{m_1}{m_2} \frac{a_2}{a_1} =
\frac{m_1}{m_2} \left( \frac{q+1}{q} \right)^{2/3},
\label{ratiotim}
\end{equation}

\noindent where we have assumed that the eccentricities are very small
compared to $H/r$.  Convergent migration, which is required for the
resonance to be maintained, corresponds to $t_{a,2}/t_{a,1} <1$.

\begin{figure}
\begin{center}
\includegraphics[scale=0.5]{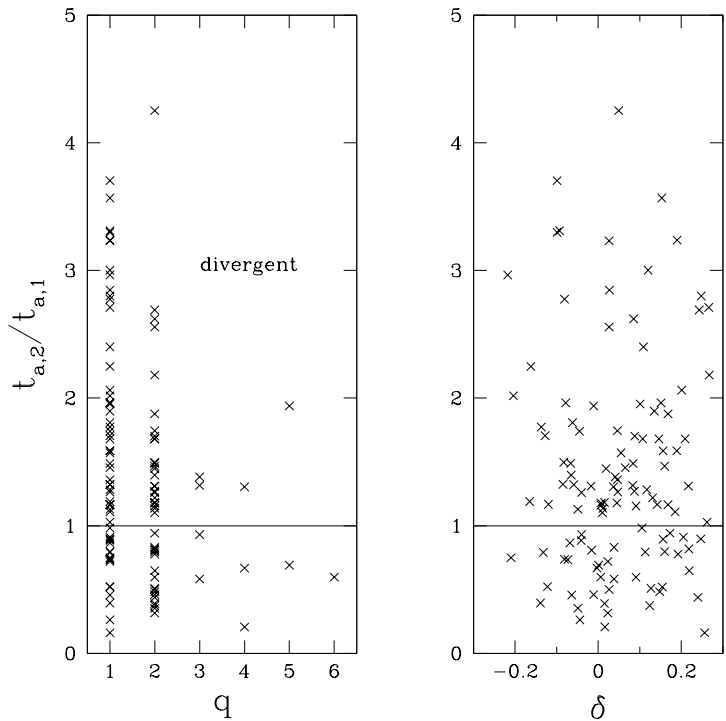}
\end{center}
\caption{ Ratio of the migration timescales $t_{a,2}/t_{a,1}$ {\em
    versus} $q$ (left--hand) and {\em versus} $\delta$ (right--hand).
  Here $q$ is chosen according to criterion~(a), but very similar
  plots would have been obtained by choosing criterion~(b) instead.
  Divergent (convergent) migration corresponds to $t_{a,2}/t_{a,1} >1$
  ($t_{a,2}/t_{a,1} <1$). Only values of $t_{a,2}/t_{a,1}$ lower than
  5 are displayed. We see that for about 65\% of the systems,
  $t_{a,2}/t_{a,1} >1$, so that MMRs cannot have been established
  during smooth migration.  Also, contrary to what could have been
  expected, there is no tendency for systems closer to resonance to
  have preferentially $t_{a,2}/t_{a,1} <1$. }
\label{fig6}
\end{figure}

\noindent Figure~\ref{fig6} shows $t_{a,2}/t_{a,1}$ as a function of
$q$ for all the systems in our sample, with $q$ chosen according to
criterion~(a) (very similar plot would have been obtained by choosing
criterion~(b) instead).  If the resonance were established and
maintained during migration, then $t_{a,2}/t_{a,1}$ given by
equation~(\ref{ratiotim}) would be smaller than 1.  We see that this
is not the case for 75 of the systems, which represents about 65\% of
the systems.  Therefore, at least in the context of our disc model,
and assuming constant planet masses, resonances for these systems have
not been established during smooth migration.

%\begin{figure}
%\begin{center}
%\includegraphics[scale=0.7]{figure5b.pdf}
%\end{center}
%\caption{ Ratio of the migration timescales $t_{a,2}/t_{a,1}$ {\em
%    versus} $q$.  For the left--hand plots, which are labelled (a),
% the value of $q$ is chosen in such a way as to minimize
%  $\left| \delta \right|$. Labels~(a) and~(b) have the same meaning as
%  in figure~\ref{fig5a}.  Divergent (convergent) migration corresponds to
%  $t_{a,2}/t_{a,1} >1$ ($t_{a,2}/t_{a,1} <1$). Only values of
%  $t_{a,2}/t_{a,1}$ lower than 5 are displayed.  }
%\label{fig5b}
%\end{figure}

\noindent Figure~\ref{fig6} also shows $t_{a,2}/t_{a,1}$ as a function of
$\delta$ for all the systems in our sample, with $q$ chosen according
to criterion~(a) (again, very similar plot would have been obtained by
choosing criterion~(b) instead).   We could have expected systems
closer to resonances, i.e. with smaller values of $\delta$, to have
preferentially $t_{a,2}/t_{a,1} <1$, which corresponds to convergent
migration, but this is not the case.  
%This is in contrast to the claim by Pan \& Schlichting (2017).
%Submitted in 2017 but not yet accepted...
Including the 5:3 resonance would not change the conclusions of this
subsection. 

\section{Evolution of a resonant system entering a cavity}
\label{sec:cavity}

%Planets are too close to resonances for that to be 'by chance', but
%too far away for a gentle process to have disrupted it.  

%2 planets move in together, the inner planet is less massive:
%convergent migration.  There is already a more massive planet that has
%moved in (more massive: goes faster), it is in the cavity.  The outer
%2 join, no disruption at first as the disc damps eccentricities.  But
%the disc dissipates.  We assumed the inner planet in the pair
%collides with the more massive planet in the cavity: the resonance is
%disrupted. 

MMRs involving only two planets are very stable, which means that once
established they are very difficult to disrupt.  It has been proposed
that chains of resonances could be disrupted when the disc dissipates,
as then eccentricities grow (Izidoro et al. 2017).  However, this is not the
case for systems comprising two planets only in the mass regime we are
investigating here.  We have tested this hypothesis by solving the set
of equations~(\ref{dvarpi1dt})--(\ref{de2dtm}) and removing the disc
adopting various timescales, but have found that in almost all cases
the resonance survives.

It has been noted from previous simulations (eg. Xiang--Gruess \&
Papaloizou~2015) that significant increases in orbital eccentricities
may occur when planets enter a cavity interior to the disc.  This is
potentially disruptive.  However, the dynamics has not been studied in
detail and is the focus of this section.  This is important because
pairs that originally had divergent migration in the smooth disc can
form MMRs after the inner planet enters a cavity.

\subsection{Disruption of the resonance when entering the cavity}

\begin{figure}
\begin{center}
\includegraphics[scale=0.7]{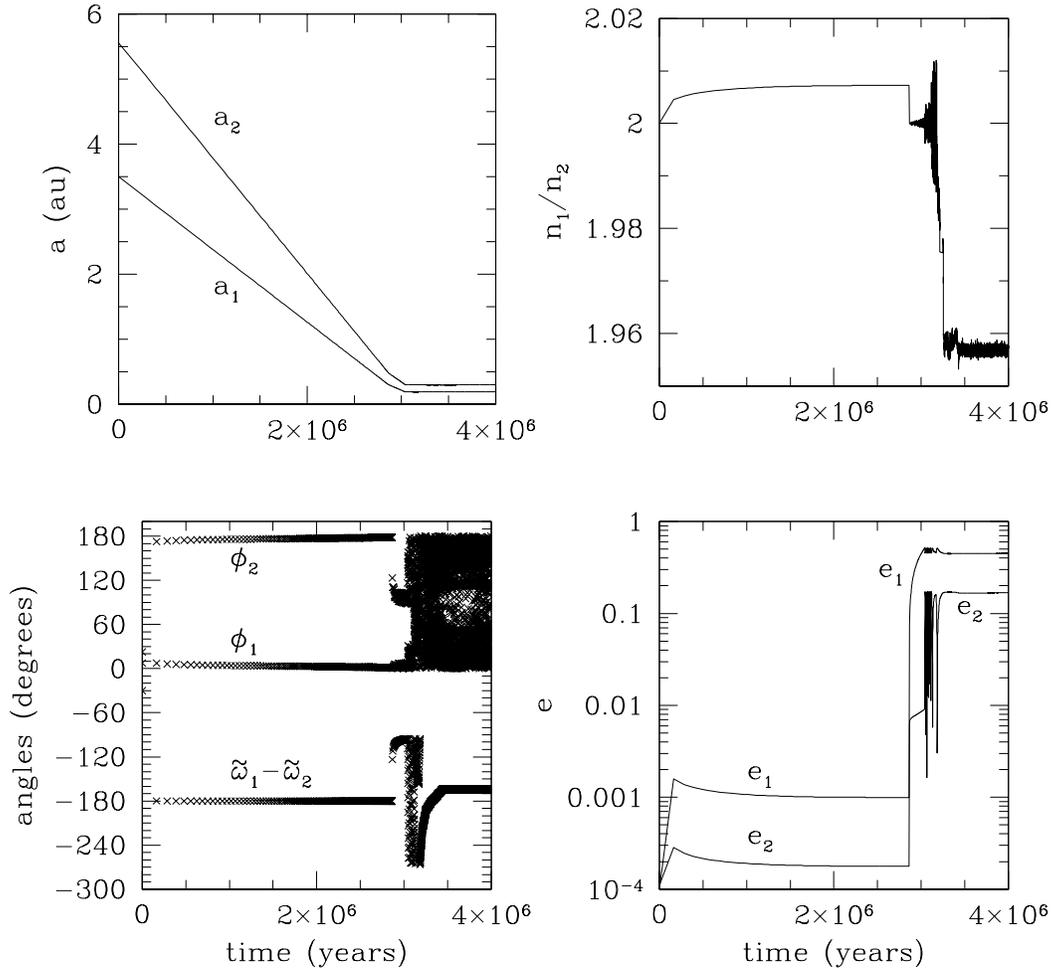}
\end{center}
\caption{Evolution of the system for $q=1$, $m_1= 1$~$\me$,
  $m_2=1.6$~$\me$, $m_d=3 \times 10^{-5}$~$\msun$ and $\rcav =0.3$~au
  as a function of time (in years).  The upper left--hand plot shows
  the semimajor axes (in au).  The upper right--hand plot shows
  $n_1/n_2$.  The lower left--hand plot shows the resonant angles
  $\phi_1$ and $\phi_2$ and the difference of the pericenter
  longitudes $\varpi_1-\varpi_2$ (in degrees).  The lower right--hand
  plot shows the eccentricities.  The resonance is disrupted when the
  inner planet enters the cavity.  Note that the calculations shown
  here cannot be trusted after the planet has entered the cavity, as
  the equations which are solved are only valid to first order in
  eccentricities.  However, disruption of the resonance may be real.  }
\label{fig7}
\end{figure}

Here we model the cavity as a discontinuity, meaning that the damping
terms are discontinuously set to zero when the planet is located
inside the inner edge.  When the innermost planet enters the inner
cavity, the resonance may be disrupted.  
This is
illustrated in figure~\ref{fig7} 
%and~\ref{fig8} 
for $q=1$.  For fixed values of $m_1$, $m_2$ and $m_d$, this happens
when the radius of the inner cavity, $\rcav$, is larger than some
critical value $\rc$.  
%This is illustrated in figures~\ref{fig7} and~\ref{fig8} for $q=1$.  
For a given $m_1$, $\rc$ decreases when $m_2$ and $m_d$ increase.  It
can be seen on  figure~\ref{fig7}   that the eccentricity $e_1$ of the
innermost planet reaches very high values after the planet enters the
cavity, as there is no longer damping from the disc.  This tends to
disrupt the MMR.  For a given MMR, the larger the cavity, the larger
the separation between the planets, and the easier it is for the MMR
to be disrupted when $e_1$ becomes large.

%\begin{figure}
%\begin{center}
%\includegraphics[scale=0.7]{figure8.pdf}
%\end{center}
%\caption{Evolution of the system for $q=1$ and $m_1= 1$~$\me$
%as a function of time (in years).  The upper left--hand plot shows
%$n_1/n_2$ for $m_2=1.6$~$\me$, $m_d=3 \times 10^{-5}$~$\msun$ and
%$\rcav=0.25$, 0.27, 0.3 and 0.4~au.   The upper right--hand plot shows
%  $n_1/n_2$ for $m_d=3 \times 10^{-5}$~$\msun$, $\rcav =0.2$~au and
%  $m_2=1.6$ and 1.8~$\me$.  The lower plots show $n_1/n_2$
%  (left--hand) and $e_1$ (right--hand) for $m_2=1.6$~$\me$, $\rcav
%  =0.05$~au and $m_d=3 \times 10^{-5}$ and $7 \times 10^{-5}$~$\msun$.
%}
%\label{fig8}
%\end{figure}

The calculations shown in figure~\ref{fig7} have been done by solving
the set of equations~(\ref{dvarpi1dt})--(\ref{de2dtm}), which are only
valid to first order in eccentricities and when the resonant angles
librate.  Therefore, although they may capture the disruption of the
resonance, they cannot be used to follow the subsequent evolution of
the system.  In the next subsection, we solve the full equations to
calculate the evolution of the system after the inner planet enters a
cavity.

\subsection{Settling into another resonance}

Here, we solve the equations of motion for each planet:

\begin{equation}
 {{\rm d}^2 {\boldsymbol{r}}_{\rm i} \over {\rm d}t^2} 
= -{G \ms
    \boldsymbol{r}_{\rm i} \over |\boldsymbol{r}_{\rm i}|^3} -
  {G m_{\rm j} \left(\boldsymbol{r}_{\rm i}- \boldsymbol{r}_{\rm j} \right) \over 
     |\boldsymbol{r}_{\rm i}- \boldsymbol{r}_{\rm j} |^3} - \sum_{k=1}^2 
     {G m_{\rm k} \boldsymbol{r}_{\rm k} 
     \over |\boldsymbol{r}_{\rm k}|^3} +
    \boldsymbol{\Gamma}_{\rm i}  \; ,
\label{emot}
\end{equation} 

\noindent where 
$\boldsymbol{r}_{\rm i}$ denotes the position vector of planet ${\rm
  i}$, and ${\rm j=2}$ or 1 for ${\rm i}=1$ or 2, respectively.  The
third term on the right--hand side is the acceleration of the
coordinate system based on the central star (indirect term).

\noindent Acceleration due to tidal interaction with the disc is dealt
with through the addition of extra forces as in Papaloizou \& Larwood
(2000, see also Terquem \& Papaloizou 2007):

\begin{equation}
\boldsymbol{\Gamma}_{\rm i} = 
-\frac{1}{\tmi} \frac{{\rm d} \boldsymbol{r}_{\rm i}}{{\rm d}t} -
\frac{2}{| \boldsymbol{r}_{\rm i}|^2 \tei} 
\left( \frac{{\rm d}  \boldsymbol{r}_{\rm i}}{{\rm d}t} \cdot
\boldsymbol{r}_{\rm i} \right) 
\boldsymbol{r}_{\rm i} ,
\end{equation}

\noindent where $\tmi=2 \tai$ and $\tei$ are the timescales on which
the angular momentum and the eccentricity of planet ${\rm i}$
decrease.  In the simulations presented below, $\tmi$ and $\tei$ are
given by equations~(\ref{tm}) and~(\ref{te}), which means that the
migration timescale in this subsection is half of what it was above.
However this does not affect our conclusions.  
  Equation~(\ref{emot}) for each planet is solved using the $N$--body
  code described in Terquem \& Papaloizou~(2007).  The evolution of
  the orbital eccentricities is accurately calculated by this code
  (see e.g., Teyssandier \& Terquem~2014 for comparisons between
  analytical and numerical results), which is important here as
  eccentricities become very large.

\begin{figure}
\begin{center}
\includegraphics[scale=0.7]{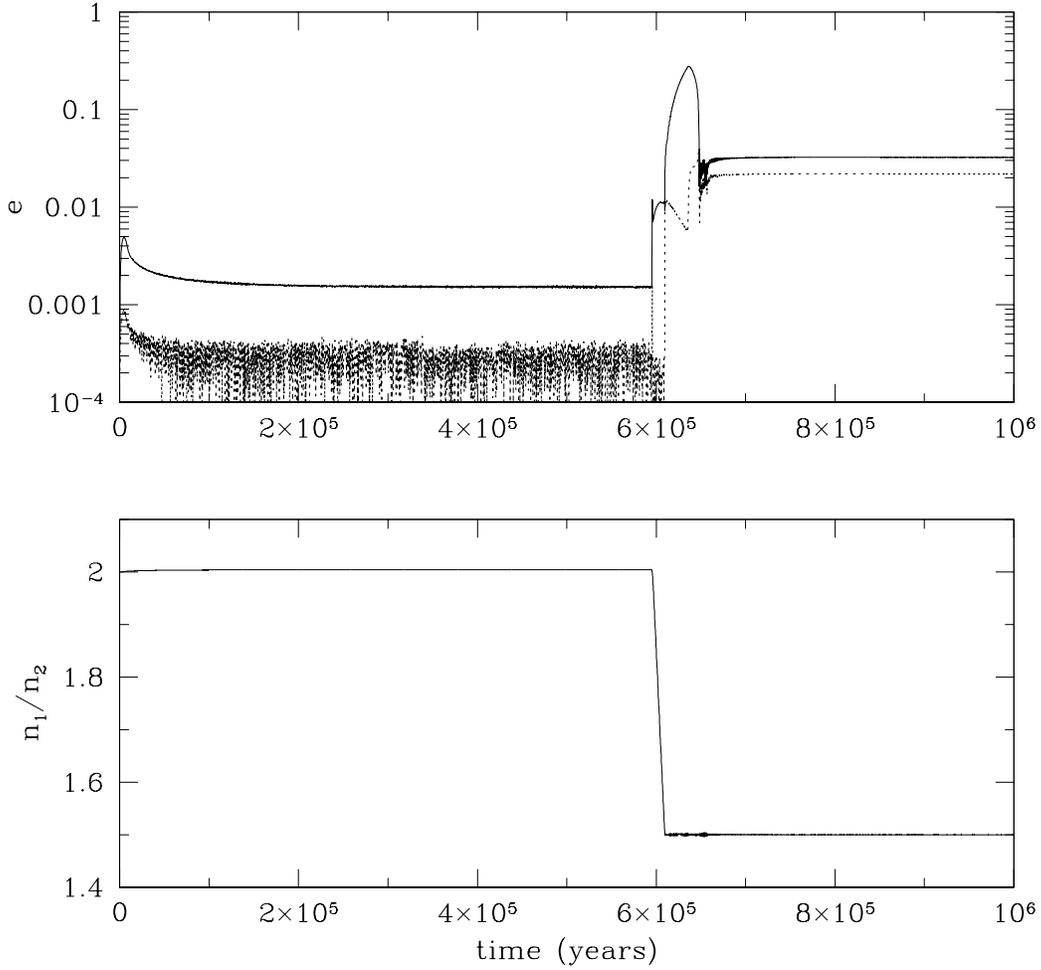}
\end{center}
\caption{
%CASE02
Evolution of the system for $q=1$, $m_1= 1$~$\me$, $m_2=1.6$~$\me$, $m_d=7 \times 10^{-5}$~$\msun$ and
$\rcav=0.4$
as a function of time (in years).  The upper  plot shows the
eccentricity of the inner planet (solid line) and that of the outer
planet (dashed line) in logarithmic scale.  The lower plot shows 
$n_1/n_2$.  The system starts in the 2:1 MMR.  When the inner planet
enters the cavity at time $t \simeq 6 \times 10^5$~years, the
resonance is disrupted.  The system subsequently evolves into the 3:2
MMR. 
}
\label{fig9}
\end{figure}

Figure~\ref{fig9} shows the evolution of a system starting
in the 2:1 MMR for
$m_1= 1$~$\me$, $m_2=1.6$~$\me$, $m_d=7 \times 10^{-5}$~$\msun$
and $\rcav=0.4$.  After the inner planet enters the cavity, its
eccentricity grows to large values and the system moves away from the
2:1 resonance.  However, it quickly settles into the 3:2 MMR.  We have
checked that the resonant angles corresponding to the 3:2 MMR librate
after that point.  In all the runs we have performed starting with
$q=1$, when the resonance is disrupted then the system moves into the
3:2 MMR.  Note that in general the 2:1 resonance is rather weak, with
the resonant angles having very large amplitude libration around
fixed values.  The 3:2 MMR is much more robust.  

\begin{figure}
\begin{center}
\includegraphics[scale=0.7]{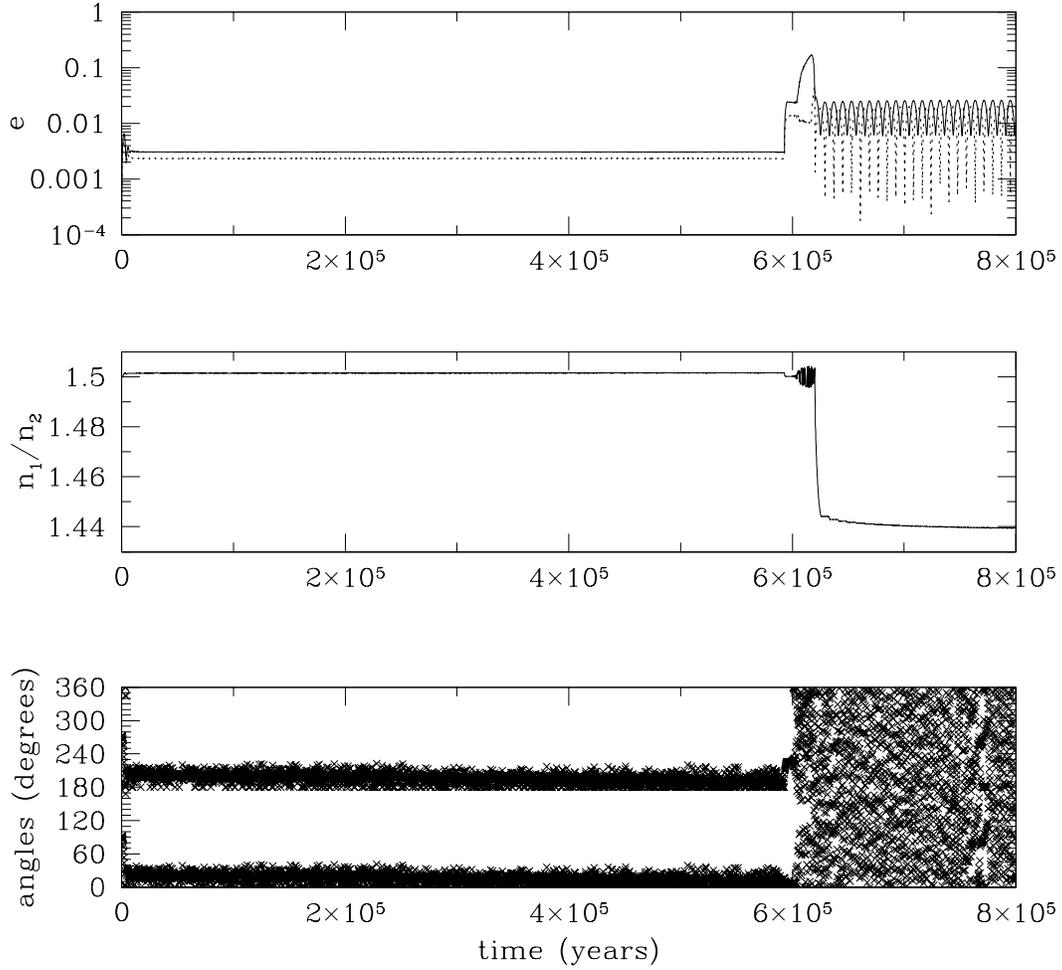}
\end{center}
\caption{
%CASE05
Evolution of the system for $q=2$, $m_1= 1$~$\me$, $m_2=1.4$~$\me$, $m_d=7 \times 10^{-5}$~$\msun$ and
$\rcav=0.3$
as a function of time (in years).  The upper  plot shows the
eccentricity of the inner planet (solid line) and that of the outer
planet (dashed line)  in logarithmic scale.  The middle plot shows 
$n_1/n_2$.  The lower plot shows the resonant angles corresponding to
the 3:2 resonance.  
The system starts in the 3:2 MMR.  When the inner planet
enters the cavity at time $t \simeq 6 \times 10^5$~years, the
resonance is disrupted, as can been seen from the fact that the
resonant angles stop librating.  The system subsequently evolves towards
$n_1/n_2=1.44$.
}
\label{fig10}
\end{figure}

Starting in exact 3:2 MMR, we have found in a number of cases that
when the MMR is disrupted the system evolves towards
$n_1/n_2 \simeq 1.45$, which means a 5\% departure from the initial
3:2 resonance.  In that case, the resonant angles corresponding to the
3:2 MMR do not librate anymore, which indicates that the resonance has
been disrupted, but this period ratio of 1.45 does not seem to
correspond to another MMR.  Therefore, there seems to be the
possibility that the resonance is disrupted and the system moves {\em
  interior} to it, but only by a few percent.  This is illustrated in
figure~\ref{fig10}.  The fact the 4:3 MMR is not reached after the
3:2 resonance is disrupted is most likely due to the fact that the
outer planet is not able to move over a distance large enough.  As can
be seen on figure~\ref{fig10}, the final period ratio of 1.45 is
attained more or less at the same time as when the eccentricities have
stabilised, which happens when the interaction with the disc ceases.
At that point, the semimajor axes are not evolving anymore.  By
contrast, in the case starting with the 2:1 MMR, the outer planet was
able to move over a distance large enough that the 3:2 MMR could be
reached.  This is supported by the fact that, in that case, the final
period ratio of 1.5 is attained before the eccentricities have
stabilised.

The large eccentricities obtained here when the inner planet enters
  the cavity may be produced in part by the fact that the damping
  timescales are set discontinuously to zero at the edge of the
  cavity.  It is possible that a smoother transition would limit the
  growth of the eccentricities.  However, the calculations above
  indicate that even with this extreme set up, the disruption of 
  resonances does not lead to systems where the two planets are
  significantly distant from a resonance.  Adopting a smoother transition
is expected to be less disruptive and so would only reinforce this conclusion.

%
%===========================================================
%

%\begin{figure}
%\begin{center}
%  \includegraphics[scale=0.7,width=0.6\textwidth,height=0.5\textwidth]
%  {TT_figure6.ps}
%\end{center}
%\caption{}
%\label{fig}
%\end{figure}

%
%===========================================================
%

\section{Discussion}
\label{sec:discussion}

In this section, we summarize our main results and discuss the
implications of our study for planet formation.  

\subsection{Summary of the main results}

In the first part of this paper, we have presented an analysis of a
first--order $(q+1):q$ resonance for any $q \ge 1$ including migration
torques.  We have derived the values of the eccentricities and
departure from exact resonance $\Delta$ at equilibrium in the case of
convergent migration.  We have also derived an expression for $\Delta$
as a function of time in the case of divergent migration.  Such an
expression had been obtained previously for small times $t$, where
$|\Delta| \propto t^{1/3}$ (Papaloizou \& Terquem ~2010, Lithwick \&
Wu~2012, Batygin \& Morbidelli~2013).  We have extended the
calculation to larger values of $t$ and incorporated the effect of
migration torques which have not previously been considered, showing
that in that regime $\Delta$ is a logarithmic function of $t$.

These analytical results have been found to be in good agreement with
the results of the numerical integration of Lagrange's planetary
equations valid to first order in eccentricities in the vicinity of
the resonance.

We have also shown that, under some circumstances, the planets could
stall interior to exact resonance.  This would happen for instance if
the planets started interior to a resonance and with no migration but
only eccentricity damping, as could be the case in parts of the disc
with appropriate mass density variations.  The system would then
expand towards exact resonance, but the separation could 
become frozen before exact resonance is reached if the planets were
resuming migration.

In the second part of the paper, we have discussed observations of
two--planet systems which are close to a resonance.  We have pointed
out that departure from exact resonance is towards larger separations
only if departures smaller than 4\% are considered.  For larger
departures, which occur for most of the systems, there is no obvious
preference for the offset to be in a particular direction. 

Finally, we have investigated the evolution of a system in a resonance
when the inner planet enters a  cavity interior to the disc.  We have found
that the 2:1 MMR is easily disrupted, but the system quickly evolves
toward the 3:2 resonance.  The 3:2 MMR is more robust, although in
some cases we have found that the period ratio decreases by a few
percent while the resonance angles stop librating.

\subsection{Migration {\em versus} in--situ formation}

The analysis of the data suggests that even when a system is very
close to MMR, the resonance in most cases cannot have been established
while the planets were migrating smoothly through the disc.
Therefore, {\em if capture in resonance does occur, it is in general
  after the planets have reached the disc's inner edge}.  That happens
if one planet migrates first and penetrates inside a cavity interior
to the disc, or stalls just beyond, and another planet subsequently
migrates down towards the cavity and locks the inner planet into a
resonance.  This scenario can explain systems close to MMRs.  However,
if migration is a general outcome and happens in all the systems,
since most systems show significant departure from exact MMR, there
has to be a process capable of disrupting significantly the resonance
after it is established in the way described above.  Alternatively,
there has to be a process that prevents the resonance from being
established.  Failing that, we have to conclude that the two planets
have formed not too far away from the disc's inner parts and that
migration has been limited, in a scenario approximating {\em in--situ}
formation.

Permanent capture into a resonance can be avoided for a range of
parameters for which the resonance is overstable, as shown by
Goldreich \& Schlichting (2014).  However, this requires the outer
planet to be more massive than the inner one (Deck \& Batygin~2015),
which is in general not the case, as discussed above.

A number of processes capable of significantly moving systems away
from resonances (by more than a few percent) have been proposed, but
so far none of them seem to be able to single--handedly explain the
data:
\begin{itemize}
\item[(i)] turbulent fluctuations in the disc can destabilize
  resonances (Adams et al. 2008). It has been shown that, with
 an  appropriate level of turbulence in the disc,  stochastic
  migration is able to produce systems with orbital parameters which are in
  agreement with the data (Rein~2012).  However, Batygin \&
  Adams~(2017) have recently argued that for any realistic parameters
  describing the disc, this process is only efficient if the total
  mass in the system if more than 3~$\me$ which, given the
  uncertainty on the masses, makes it rather marginal though possibly
  not working for the largest masses.   
\item[(ii)] interaction between a planet and the wake of a companion
  produces significant departure from exact resonance
  (Baruteau \& Papaloizou 2013).  Note that  this process does not
  necessarily require the MMR to be established through smooth
  convergent migration, although that was the case investigated by
  Baruteau \& Papaloizou (2013).  It would also work if the MMR were
  established with the inner planet near a cavity edge where the
  surface density decreased smoothly inwards, and in that case the
  outer planet may not have to be more massive.  However, this process
  requires a particular relation between the planet masses and disc
  properties to work, so it is unlikely to be universal. 
\item[(iii)] departure from exact resonance may be significant if capture
  into resonance occurs during migration in a flared disc (Ramos et
  al. 2017).  However,  this would require convergent migration of all
  the systems near resonance, which is not consistent with the data.
\item[(iv)] planets could move out of resonance after reaching the
  disc inner parts if the magnetospheric cavity expands and planets
  are trapped beyond the edge of the cavity and move outward with it
  (Liu et al. 2017).  However, departure from resonance is induced
  only when the outer planet is more massive than the inner one, which
  is not generally the case, as shown above.
\item[(v)] resonances may be disrupted when the disc dissipates, as
  eccentricities get excited to high values (Izidoro et al. 2017).
  This requires more than two planets in the system (we have
    checked that resonances do survive disc dissipation for a broad
    range of parameters when only two planets are involved in the
    resonance) but, as inclinations are produced through this
    dynamical instability, transit observations may lead to only two
    planets being detected.  However, even though the planets in
the calculations of   Izidoro et al.~(2007)  are significantly more massive
    than those in the Kepler sample,  the fraction
    of stable resonant chains obtained in their model  is significantly higher than that  needed
    to match the distribution of observed planets. 
\item[(vi)] departures from resonance may happen after the gas in the
  disc dissipates and as a result of interactions between the planets
  and planetesimals (Chatterjee \& Ford~2015).  However, for
  significant departure to occur, the mass in the planetesimal disc
  has to be at least half the mass of the planets themselves.  Such
  massive planetesimal populations would be unlikely in the inner
  parts of discs. 
\end{itemize}

None of these processes taken in isolation can explain the range of
observations, and it has yet to be shown whether when taken together
they can reproduce the spread of period ratios which is observed.
Strict {\em in--situ} formation of low mass planets has been
investigated in previous studies.  Hansen \& Murray (2013) have shown
that the output of their Monte Carlo model for the structure of low
mass planets that form {\em in--situ} is in rather good agreement with
{\em Kepler} observations, except for the fact that it does not
produce enough single planet systems.  Petrovich et al. (2013) have
found using a simplified model that the distribution of period ratios
for planets forming {\em si--situ} is similar to that observed by {\em
  Kepler}, which peaks around resonances.  However, their model
produces planets with final masses significantly exceeding those of
super--Earths.
% and with large orbital eccentricities.  
Note that, in the {\em in--situ} formation scenario, for migration and
resonant capture to be avoided, planets have to finish growing after
most of the gas has been depleted.

Strict {\em in--situ} formation has not yet been shown convincingly to
be able to explain the data.  In addition, the existence of resonant
chains indicates that some migration does occur.  However, as shown in
this paper, there is no support from the observations for extensive
(over a large radial extent) convergent migration in a smooth disc.
Only a small fraction of the systems have migrated through the disc
and established a MMR (either during migration or after reaching the
disc's inner parts).  

The above discussion suggests that there may be two populations of low
mass planets:  
\begin{itemize}
\item[(1)] A small population where smooth migration was extensive so
  that MMRs were readily produced in the extended disc when it was
  convergent or near the cavity when it was not.  In these systems,
  the planets have subsequently separated slightly, possibly due to
  tidal interaction with the star or other dissipative process.  
    If we assume that all the systems in our sample with
    $0 \le \delta < 0.04$ belong to this population, then the fraction of
    systems in this population is about 15\%.
\item[(2)]Another larger population for which migration was much more
  modest, producing MMRs only in a small number of cases, this
  approximating {\em in--situ} formation.  
\end{itemize}

Which scenario prevails may depend on the initial disc's mass.
Terquem (2017) pointed out that in low--mass discs, cores forming at
around 1~au or beyond do not have enough time to migrate down to the
disc's inner parts.  This is because the disc photoevaporates before
migration of these cores can become significant.  If systems close to
the star and which have significant departure from MMRs have formed
approximately {\em in--situ}, migration was not efficient in the disc
in which they formed and therefore we would expect more low mass
planets to be present further away.

%\section*{Acknowledgements}

%
%===========================================================
%

%\bibliographystyle{apj}
%\bibliographystyle{plain}
%\bibliographystyle{mn2e}
%\bibliography{biblio_papers}

%
%===========================================================
%

\appendix

\section{Coefficients in the disturbing function}

\label{app:coef}

In table~\ref{tab:coeffs}, we give the values of the coefficients $f_1$
and $f_2$, calculated from equations~(\ref{f1}) and~(\ref{f2}), for
$q$ between 1 and 6.  Since
$f_2^{\prime} \equiv f_2 - 2 \alpha \delta_{q,1}$, we have
$f_2^{\prime} = f_2 - 2^{1/3}$ for $q=1$ and $f_2^{\prime} = f_2$ for
$q>1$.

\begin{table}
\begin{tabular}{ccc}
  \hline
  \hline
  $q$ & $f_1$ & $f_2$ \\
  \hline
1 & $-1.190$ & 1.688 \\
2 & $-2.025$ & 2.484 \\
3 & $-2.840$ & 3.283 \\
4 & $-3.650$ & 4.084 \\
5 & $-4.456$ & 4.885 \\
6 & $-5.261$ & 5.686 \\
  \hline
  \hline
\end{tabular}
\label{tab:coeffs}
\caption{Numerical values of $f_1$ and $f_2$ for $q$ between 1 and 6.}
\end{table}

%===

\end{document}